%% file: geom-mean.tex
\documentclass[11pt,a4paper,english]{article}
\pdfoutput=1

\usepackage{lmodern}
\renewcommand{\sfdefault}{lmss}

\usepackage{color}
\usepackage{float}
\usepackage{mathtools}
\usepackage{amsmath}
\usepackage{amsthm}
\usepackage{amssymb}
\usepackage{stmaryrd}
\usepackage{graphicx}

\makeatletter

\providecommand{\tabularnewline}{\\}

\input{preamble.tex}

\makeatother

\usepackage{babel}
\begin{document}

\input{macros.tex}

%%%%%%%%%%%%%%%%%%%%%%%%%%%%%%%%%%%%%%%%%%%%%%%%%%%%%%%%%%%%%%%%%%%%%%%%%%%%%
% Front matter
%%%%%%%%%%%%%%%%%%%%%%%%%%%%%%%%%%%%%%%%%%%%%%%%%%%%%%%%%%%%%%%%%%%%%%%%%%%%%

\newcommand{\myReleaseInfo}{~\\~}

\newcommand{\myAbstract}{We use the geometric mean to parametrize
metrics in the Hassan\textendash Rosen ghost-free bimetric theory
and pose the initial-value problem. The geometric mean of two positive
definite symmetric matrices is a well-established mathematical notion
which can be under certain conditions extended to quadratic forms
having the Lorentzian signature, say metrics $g$ and $f$. In such
a case, the null cone of the geometric mean metric $h$ is in the
middle of the null cones of $g$ and $f$ appearing as a geometric
average of a bimetric spacetime. The parametrization based on $h$
ensures the reality of the square root in the ghost-free bimetric
interaction potential. Subsequently, we derive the standard $n$+1
decomposition in a frame adapted to the geometric mean and state the
initial-value problem, that is, the evolution equations, the constraints,
and the preservation of the constraints equation.}

\newcommand{\myTitle}{Geometric mean of bimetric spacetimes}

\newcommand{\myKeywords}{\bgroup\small Modified gravity, Ghost-free
bimetric theory, Geometric mean, Initial-value problem\egroup}

\title{\myTitle}

\author{Mikica Kocic}

\affiliation{
  Department of Physics \& The Oskar Klein Centre,\\
  Stockholm University, AlbaNova University Centre,
  SE-106 91 Stockholm
}

\email{mikica.kocic@fysik.su.se}

\hypersetup{
  pdftitle=\myTitle,
  pdfauthor=Mikica Kocic,
  pdfsubject=Hassan-Rosen ghost-free bimetric theory,
  pdfkeywords={Modified gravity, Ghost-free bimetric theory, 
    Geometric mean, Initial-value problem}
}

%\notoctrue
\emitFrontMatter
\vspace{2mm}

%%%%%%%%%%%%%%%%%%%%%%%%%%%%%%%%%%%%%%%%%%%%%%%%%%%%%%%%%%%%%%%%%%%%%%%%%%%%%
% Main matter
%%%%%%%%%%%%%%%%%%%%%%%%%%%%%%%%%%%%%%%%%%%%%%%%%%%%%%%%%%%%%%%%%%%%%%%%%%%%%

\input{sec-10.tex}

\input{sec-20.tex}

\input{sec-30.tex}

\input{sec-40.tex}

\input{sec-60.tex}

%%%%%%%%%%%%%%%%%%%%%%%%%%%%%%%%%%%%%%%%%%%%%%%%%%%%%%%%%%%%%%%%%%%%%%%%%%%%%
% Appendices
%%%%%%%%%%%%%%%%%%%%%%%%%%%%%%%%%%%%%%%%%%%%%%%%%%%%%%%%%%%%%%%%%%%%%%%%%%%%%

\clearpage
\emitAppendix
\addtocontents{toc}{\vspace{1ex}}

\input{sec-90.tex}

\addtocontents{toc}{~}

%%%%%%%%%%%%%%%%%%%%%%%%%%%%%%%%%%%%%%%%%%%%%%%%%%%%%%%%%%%%%%%%%%%%%%%%%%%%%
% Bibliography
%%%%%%%%%%%%%%%%%%%%%%%%%%%%%%%%%%%%%%%%%%%%%%%%%%%%%%%%%%%%%%%%%%%%%%%%%%%%%

\clearpage
\ifprstyle

\bibliographystyle{apsrev4-1}
\bibliography{geom-mean}

\else

\bibliographystyle{JHEP}
\bibliography{geom-mean}

\fi

\end{document}

%% file: preamble.tex
\newif\ifprstyle

\ifx\affiliation\undefined
  \prstylefalse
\else
  \prstyletrue
  \newif\ifnotoc
  \notoctrue
\fi

\ifprstyle
  \usepackage[
    colorlinks=true, pdfa=true,
    urlcolor=blue,   anchorcolor=blue,
    citecolor=blue,  filecolor=blue,
    linkcolor=blue,  menucolor=blue,
    pagecolor=blue,  linktocpage=true
  ]{hyperref}
\else
  \usepackage{jheppub}
  \newcommand{\email}[1]{\emailAdd{#1}}
\fi

\ifx\ifCrop\undefined\else

  \ifx\ifPresentation\undefined
     \usepackage[paperwidth=171.2mm,paperheight=261.2mm]{geometry}
  \else
     \usepackage[paperwidth=330mm,paperheight=233mm]{geometry}
  \fi

  \divide\@tempdima\baselineskip
  \@tempcnta=\@tempdima
  \ifx\ifPresentation\undefined
     \hypersetup{pdfstartview={Fit},pdfpagelayout={TwoPageRight}}
  \else
    \hypersetup{pdfstartview={Fit},pdfpagelayout={SinglePage}}

    \usepackage{everypage}
    \AddEverypageHook{\tikz[remember picture,overlay]{%
      \draw [gray!10] ($(current page.east)+(-130mm,+180mm)$)
         -- ($(current page.east)+(-130mm,-100mm)$);}}
  \fi
\fi

\hypersetup{
    urlcolor=black!40!blue,   anchorcolor=black!40!blue,
    citecolor=black!40!blue,  filecolor=black!40!blue,
    linkcolor=black!40!blue,  menucolor=black!40!blue,
    pagecolor=black!40!blue
}

\usepackage{bookmark}

\usepackage{amsmath,amsfonts,amssymb,bm,cancel}

\usepackage{color,xcolor}
\usepackage{colortbl}

\usepackage{tensind}
\tensordelimiter{?}

\ifprstyle
  \newcommand{\repEqq}[2]{\tag{\ref{#1}#2}}
\else
  \newcommand{\repEqq}[2]{\tag{\ref*{#1}#2}}
\fi

\newcommand{\bSe}{\begin{subequations}}
\newcommand{\eSe}{\end{subequations}}

\newcommand{\bWe}{\begin{widetext}}
\newcommand{\eWe}{\end{widetext}}

\newtheorem{proposition}{Proposition}
\newtheorem{lemma}{Lemma}

\allowdisplaybreaks

\usepackage{tikz}
\usetikzlibrary{shapes,arrows}
\usetikzlibrary{calc}
\usetikzlibrary{positioning}
\usetikzlibrary{decorations.pathreplacing}
\usetikzlibrary{shapes.misc}

\usepackage[scr,scaled=1.1]{rsfso}
\usepackage{eucal}

\DeclareMathAlphabet{\mathsfit}{\encodingdefault}{\sfdefault}{m}{sl}
\SetMathAlphabet{\mathsfit}{bold}{\encodingdefault}{\sfdefault}{bx}{sl}

\usepackage{float}

\floatstyle{plaintop}
\newfloat{bbox}{thp}{lop}
\floatname{bbox}{Box}

\renewcommand\floatc@plain[2]{\setbox\@tempboxa\hbox{{\@fs@cfont #1.} #2}%
\ifdim\wd\@tempboxa>\hsize {\@fs@cfont #1.} #2\par
\else\hbox to\hsize{\hfil\box\@tempboxa\hfil}\fi}

\ifprstyle\else
  \usepackage[toc]{multitoc}
  
  \addtocontents{toc}{\protect\small}
\fi

\newcommand{\emitFrontMatter}{
  \ifprstyle
    \begin{abstract}\myAbstract\end{abstract}
  \else
    \ifnotoc
      \abstract{\\[-1.64ex]\hphantom{\hspace{0.28cm}}%
        \parbox{1\columnwidth-0.28cm}%
        {\renewcommand\baselinestretch{1.05}\small\hspace{18mm}\myAbstract}}
    \else
      \abstract{\myAbstract}
    \fi
  \fi

  \keywords{\myKeywords}

  \ifprstyle\else
    \makeatletter
      \def\@fpheader{\myReleaseInfo}
    \makeatother
    \ifnotoc
      \compress
      \renewcommand\afterLogoSpace{}
      \renewcommand\afterSubheaderSpace{}
      \renewcommand\afterProceedingsSpace{}
      \makeatletter
      \renewcommand\ps@titlepage{}
      \makeatother
      \toccontinuoustrue
    \else
      \renewcommand\afterTocSpace{\medskip}
      \addtocontents{toc}{\protect\setcounter{tocdepth}{2}}
    \fi
  \fi

  \maketitle

  \ifprstyle\else
    \ifnotoc
      \hrule\bigskip\bigskip
    \else
      \flushbottom
    \fi
  \fi
}

\usepackage{titlesec}

\newcommand{\emitAppendix}{
  \phantomsection
  \addcontentsline{toc}{section}{Appendices}
  \addtocontents{toc}{\protect\setcounter{tocdepth}{2}}

  \makeatletter
    \def\toclevel@section{1}
    \def\toclevel@subsection{2}
  \makeatother

  \addtocontents{toc}{\protect\makeatletter}
  \addtocontents{toc}{\string\let\string\l@chapter\string\l@section}
  \addtocontents{toc}{\string\let\string\l@section\string\l@subsection}
  \addtocontents{toc}{\protect\makeatother}

  \titleformat{\section}{\normalfont\large\bfseries}{Appendix~\thesection~~~}{0em}{}

  \let\oldSection\section
  \renewcommand\section{\bookmarksetupnext{level=2}\oldSection}

  \appendix
}

%% file: macros.tex
\newif\ifColors

\newif\ifShowSigns 

%%%%%%%%%%%%%%%%%%%%%%%%%%%%%%%%%%%%%%%%%%%%%%%%%%%%%%%%%%%%%%%%%%%%%%%%%%%%%
% Avoid the redefinitions of the commands if already defined
\ifx \ii \undefined

% Adjusted color flavors
\definecolor{red}{rgb}{1,0,0.1}
\definecolor{green}{rgb}{0.0,0.6,0}
\definecolor{blue}{rgb}{0.1,0.1,1}
\definecolor{orange}{rgb}{0.6,0.3,0}
\definecolor{magenta}{rgb}{0.9,0.1,1}

\newcommand\nPlusOne{$N$+1}

\renewcommand\tilde[1]{\mkern1mu\widetilde{\mkern-1mu#1}}

\global\long\def\ii{\mathrm{i}}
\global\long\def\ee{\mathrm{e}}
\global\long\def\dd{\mathrm{d}}
\global\long\def\ppi{\mathrm{\pi}}
\global\long\def\tr{\mathsf{{\scriptscriptstyle T}}}
\global\long\def\Tr{\operatorname{Tr}}
\global\long\def\op#1{\operatorname{#1}}
\global\long\def\dim{\operatorname{dim}}
\global\long\def\diag{\operatorname{diag}}
\global\long\def\Lie{\mathrm{\mathscr{L}}}

\global\long\def\mfrac#1#2{\frac{\raisebox{-0.45ex}{\scalebox{0.9}{#1}}}{\raisebox{0.4ex}{\scalebox{0.9}{#2}}}}
\global\long\def\mbinom#1#2{\Big(\begin{array}{c}
 #1\\[-0.75ex]
 #2 
\end{array}\Big)}

\global\long\def\QED{\mbox{{\color{red}\qedhere}}}

\global\long\def\tudu#1#2#3#4{?{\mbox{\ensuremath{#1}}}^{#2}{}_{#3}{}^{#4}?}
\global\long\def\tdud#1#2#3#4{?{\mbox{\ensuremath{#1}}}{}_{#2}{}^{#3}{}_{#4}?}
\global\long\def\tud#1#2#3{?{\mbox{\ensuremath{#1}}}^{#2}{}_{#3}?}
\global\long\def\tdu#1#2#3{?{\mbox{\ensuremath{#1}}}{}_{#2}{}^{#3}?}

\global\long\def\repEq#1#2{\repEqq{#1}{#2}}

\global\long\def\Aeq#1{\tag*{(#1)}}
\global\long\def\PSeq#1{\tag*{\llap{P\,}(#1)}}
\global\long\def\BJeq#1{\tag*{\llap{B\,}(#1)}}

\global\long\def\qvf{\xi}
\global\long\def\ixA{a}
\global\long\def\ixB{b}
\global\long\def\ccVar{\mathcal{C}}

\global\long\def\lidx#1{\ ^{(#1)}\!}

\global\long\def\gSector#1{{#1}}
\global\long\def\fSector#1{{#1}}
\global\long\def\hSector#1{{#1}}
\global\long\def\sSector#1{{#1}}
\global\long\def\mSector#1{{#1}}
\global\long\def\hrColor#1{{#1}}
\global\long\def\VColor#1{{#1}}
\global\long\def\KColor#1{{#1}}
\global\long\def\KVColor#1{{#1}}

\global\long\def\gMet{\gSector g}
\global\long\def\gSp{\gSector{\gamma}}
\global\long\def\gLapse{\gSector N}
\global\long\def\gShift{\gSector N}
\global\long\def\gShiftVec{\gSector{\vec{N}}}
\global\long\def\gK{\gSector K}
\global\long\def\gE{\gSector e}
\global\long\def\gD{\gSector D}
\global\long\def\gR{\gSector R}
\global\long\def\gCS{\gSector{\Gamma}}
\global\long\def\gVse{\gSector{V_{g}}}
\global\long\def\gTse{\gSector{T_{g}}}
\global\long\def\gEinst{\gSector{G_{g}}}
\global\long\def\gRicci{\gSector{R_{g}}}
\global\long\def\gCC{\gSector{\mathcal{C}}}
\global\long\def\gCE{\gSector{\mathcal{E}}}
\global\long\def\gCD{\gSector{\nabla}}

\global\long\def\fMet{\fSector f}
\global\long\def\fSp{\fSector{\varphi}}
\global\long\def\fLapse{\fSector M}
\global\long\def\fShift{\fSector M}
\global\long\def\fShiftVec{\fSector{\vec{M}}}
\global\long\def\fK{\fSector{\tilde{K}}}
\global\long\def\fE{\fSector m}
\global\long\def\fD{\fSector{\tilde{D}}}
\global\long\def\fR{\fSector{\tilde{R}}}
\global\long\def\fCS{\fSector{\tilde{\Gamma}}}
\global\long\def\fVse{\fSector{V_{f}}}
\global\long\def\fTse{\fSector{T_{f}}}
\global\long\def\fEinst{\fSector{G_{f}}}
\global\long\def\fRicci{\fSector{R_{f}}}
\global\long\def\fCC{\fSector{\widetilde{\mathcal{C}}}}
\global\long\def\fCE{\fSector{\widetilde{\mathcal{E}}}}
\global\long\def\fCD{\fSector{\widetilde{\nabla}}}

\global\long\def\gKappa{\gSector{\kappa_{g}}}
\global\long\def\gKappainv{\gSector{\kappa_{g}^{-1}}}
\global\long\def\Mg{\gSector{M_{g}^{d-2}}}

\global\long\def\fKappa{\fSector{\kappa_{f}}}
\global\long\def\fKappainv{\fSector{\kappa_{f}^{-1}}}
\global\long\def\Mf{\fSector{M_{f}^{d-2}}}

\global\long\def\grho{\gSector{\rho}}
\global\long\def\gjota{\gSector j}
\global\long\def\gJota{\gSector J}

\global\long\def\frho{\fSector{\tilde{\rho}}}
\global\long\def\fjota{\fSector{\tilde{j}}}
\global\long\def\fJota{\fSector{\tilde{J}}}

\global\long\def\gAlpha{\gSector{\alpha}}
\global\long\def\gBeta{\gSector{\beta}}
\global\long\def\gEA{\gSector A}
\global\long\def\gEB{\gSector B}
\global\long\def\fAlpha{\fSector{\tilde{\alpha}}}
\global\long\def\fBeta{\fSector{\tilde{\beta}}}
\global\long\def\fEA{\fSector{\tilde{A}}}
\global\long\def\fEB{\fSector{\tilde{B}}}

\global\long\def\sEtau{\mSector{\tau}}
\global\long\def\sESigma{\mSector{\Sigma}}
\global\long\def\sER{\mSector R}

\global\long\def\Proj{\operatorname{\perp}}
\global\long\def\gProj{\gSector{\operatorname{\perp}_{g}}}
\global\long\def\fProj{\fSector{\operatorname{\perp}_{f}}}
\global\long\def\hProj{\hSector{\operatorname{\perp}}}
\global\long\def\prho{\boldsymbol{\rho}}
\global\long\def\pjota{\boldsymbol{j}}
\global\long\def\pJota{\boldsymbol{J}}

\global\long\def\sgn{\gSector{\mathsfit{n}{\mkern1mu}}}
\global\long\def\sgD{\gSector{\mathcal{D}}}
\global\long\def\sgQ{\gSector{\mathcal{Q}}}
\global\long\def\sgV{\gSector{\mathcal{V}}}
\global\long\def\sgU{\gSector{\mathcal{U}}}
\global\long\def\sgB{\gSector{\mathcal{B}}}

\global\long\def\sfn{\fSector{\tilde{\mathsfit{n}}{\mkern1mu}}}
\global\long\def\sfD{\fSector{\widetilde{\mathcal{D}}}}
\global\long\def\sfQ{\fSector{\widetilde{\mathcal{Q}}}}
\global\long\def\sfV{\fSector{\widetilde{\mathcal{V}}}}
\global\long\def\sfU{\fSector{\widetilde{\mathcal{U}}}}
\global\long\def\sfB{\fSector{\widetilde{\mathcal{B}}}}

\global\long\def\sgW{\gSector{\mathcal{W}}}
\global\long\def\sgQU{\gSector{(\mathcal{Q\fSector{{\scriptstyle \widetilde{U}}}})}}

\global\long\def\sfW{\fSector{\tilde{\mathcal{W}}}}
\global\long\def\sfQU{\fSector{(\mathcal{\widetilde{Q}\gSector{{\scriptstyle U}}})}}

\global\long\def\hMet{\hSector h}
\global\long\def\hSp{\hSector{\chi}}
\global\long\def\hLapse{\hSector H}
\global\long\def\hShift{\hSector q}
\global\long\def\hShiftVec{\hSector q}
\global\long\def\hCC{\hSector{\bar{\mathcal{C}}}}

\global\long\def\sLs{\sSector{\hat{\Lambda}}}
\global\long\def\sLt{\sSector{\lambda}}
\global\long\def\sLtinv{\sSector{\lambda^{-1}}}
\global\long\def\sLv{\hSector v}
\global\long\def\sLp{\hSector p}
\global\long\def\sRs{\hSector{\hat{R}}}
\global\long\def\sRbar{\hSector{\bar{R}}}

\global\long\def\sI{\sSector{\hat{I}}}
\global\long\def\sEta{\sSector{\hat{\delta}}}

\global\long\def\betap#1{\beta_{{\scriptscriptstyle (#1)}}}
\global\long\def\betaScale{\ell^{-2}}
\global\long\def\betaSum{\betaScale{\textstyle \sum_{n}}\beta_{(n)}}
\global\long\def\betaSumL{\betaScale{\displaystyle \sum_{n=0}^{4}}\beta_{(n)}}

\global\long\def\signV{\,+\,}
\global\long\def\isignV{\,-\,}
\global\long\def\usignV{}
\global\long\def\uisignV{-\,}

\global\long\def\isignV{\,+\,}
\global\long\def\signV{\,-\,}
\global\long\def\uisignV{}
\global\long\def\usignV{-\,}

\global\long\def\signK{\,+\,}
\global\long\def\isignK{\,-\,}
\global\long\def\usignK{}
\global\long\def\uisignK{-\,}

\global\long\def\isignK{\,+\,}
\global\long\def\signK{\,-\,}
\global\long\def\uisignK{}
\global\long\def\usignK{-\,}

\global\long\def\signKV{\,+\,}
\global\long\def\isignKV{\,-\,}
\global\long\def\usignKV{}
\global\long\def\uisignKV{-\,}

\global\long\def\isignKV{\,+\,}
\global\long\def\signKV{\,-\,}
\global\long\def\uisignKV{}
\global\long\def\usignKV{-\,}

\global\long\def\signKV{\,+\,}
\global\long\def\isignKV{\,-\,}
\global\long\def\usignKV{+}
\global\long\def\uisignKV{-\,}

\global\long\def\isignKV{\,+\,}
\global\long\def\signKV{\,-\,}
\global\long\def\uisignKV{}
\global\long\def\usignKV{-\,}

\global\long\def\signKV{\,+\,}
\global\long\def\isignKV{\,-\,}
\global\long\def\usignKV{}
\global\long\def\uisignKV{-\,}

\global\long\def\hrD{\hrColor D}
\global\long\def\hrQ{\hrColor Q}
\global\long\def\hrn{\hrColor n}
\global\long\def\hrDn{\hrColor{Dn}}
\global\long\def\hrx{\hrColor x}

\global\long\def\hrV{\hrColor V}
\global\long\def\hrU{\hrColor U}
\global\long\def\hrVbar{\hrColor{\bar{V}}}
\global\long\def\hrWbar{\hrColor{\bar{W}}}
\global\long\def\hrSV{\hrColor S}
\global\long\def\hrUtilde{\hrColor{\tilde{U}}}
\global\long\def\hrVubar{\hrColor{\underbar{V}}}

\fi
%%%%%%%%%%%%%%%%%%%%%%%%%%%%%%%%%%%%%%%%%%%%%%%%%%%%%%%%%%%%%%%%%%%%%%%%%%%%%

%% file: sec-10.tex
\section{Introduction}

We use the geometric mean to parametrize metrics in the Hassan\textendash Rosen
(HR) ghost-free bimetric theory and pose the initial-value problem.
The HR bimetric theory \cite{Hassan:2011zd,Hassan:2011ea,Hassan:2017ugh,Hassan:2018mbl}
is a nonlinear theory of two interacting spin-2 classical fields which
is free of instabilities such as the Boulware\textendash Deser ghost
\cite{Boulware:1973my}. The HR theory is related to de Rham\textendash Gabadadze\textendash Tolley
massive gravity \cite{deRham:2010ik,deRham:2010kj,Hassan:2011hr}.
Recent reviews of these theories can be found in \cite{Schmidt-May:2015vnx,deRham:2014zqa}.

The geometric mean of two positive definite symmetric matrices  can
be written \cite{Pusz:1975aa}, 
\begin{equation}
A\op{\#}B\coloneqq A\,(A^{-1}B)^{1/2}=B\op{\#}A.
\end{equation}
Here $M^{1/2}$ denotes the principal square root of a square matrix
$M$ with no eigenvalues on the negative real axis $\mathbb{R}^{-}$,
that is, $M^{1/2}$ is the unique solution $X$ of the matrix equation
$X^{2}=M$ whose eigenvalues lie in the right half plane \cite{Higham:2008}.

The notion of the geometric mean can be extended to symmetric bilinear
forms with the Lorentzian signature under certain conditions. Given
two spacetime metrics $g$ and $f$, one can define their geometric
mean as,
\begin{equation}
h=g\op{\#}f=g\,(g^{-1}f)^{1/2}=f\op{\#}g=f\,(f^{-1}g)^{1/2}=h^{\tr},\label{eq:gm-1}
\end{equation}
provided that the principal square root $S\coloneqq(g^{-1}f)^{1/2}$
exists. According to the theorem from \cite{Hassan:2017ugh}, the
principal square root $S$ exists if and only if there exists a common
timelike direction and a common spacelike hypersurface element relative
to both $g$ and $f$. In this context, the geometric mean has a nice
geometrical interpretation: The null cone of the mean metric $h=g\op{\#}f=gS$
will be in the middle of the null cones of $g$ and $f$.

\begin{figure}
\noindent \centering{}\includegraphics[scale=0.92]{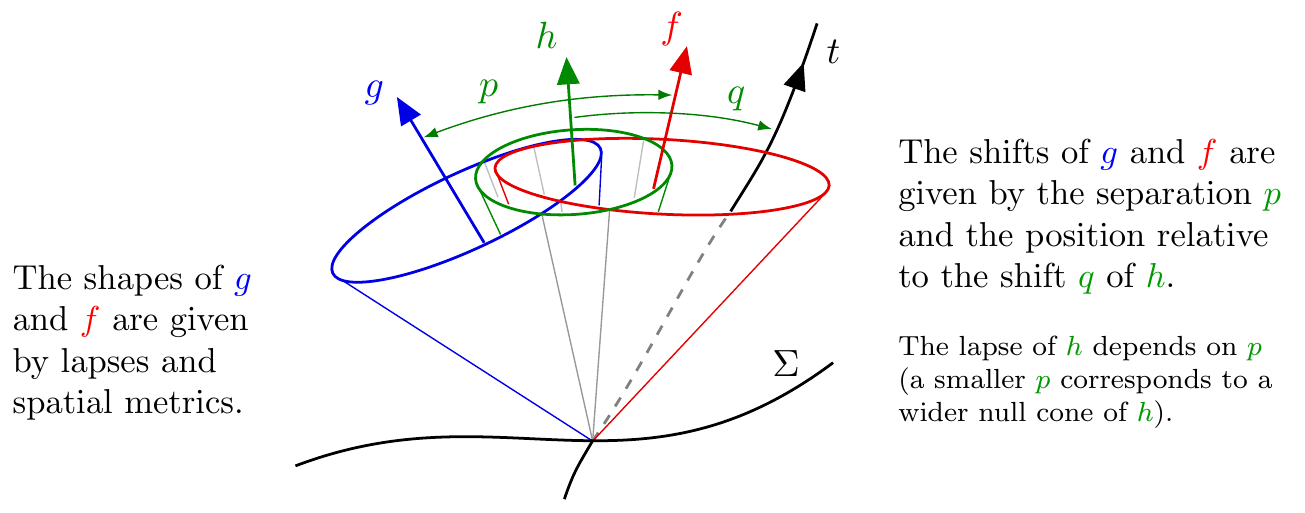}\vspace{-3ex}\caption{\label{fig:gm-param}The parametrization of $g$ and $f$ relative
to the geometric mean metric $h$.}
\end{figure}

Consequently, if a solution to the HR bimetric field equations exists,
there will always be a geometric mean metric $h$. This makes $h$
useful when adapting the space\textendash plus\textendash time foliation.
Moreover, it can be used to parametrize possible real square root
realizations of metrics with respect to $h$. Let us consider a foliation
adapted to $h$. We first specify the shift vector $q$ of $h$, as
illustrated in Figure~\ref{fig:gm-param}. Next, we split the degrees
of freedom of $g$ and $f$ across $g$, $f$ and $h$. We parametrize
the volume contributing parts of $g$ and $f$ by their lapse functions
and the spatial metrics. Finally, in addition to the slice dependent
shift $q$, we give the separation $p$ between the null cones of
$g$ and $f$ also relative to $h$. This gives all possible configurations
of $g$ and $f$ which have the real principal square root. 

After the parametrization, we decompose the bimetric field equations
and state the initial-value problem. The space-plus-time decomposition
is done for an arbitrary spacetime of dimension $d$$\,=\,$\nPlusOne{}
using the prescription from \cite{York:1979aa}. The projection yields
the constraint equations and the evolution equations in standard \nPlusOne{}
form. Furthermore, the projection of the contracted Bianchi identities
yields an additional equation that preserves the ghost-free bimetric
potential when in \nPlusOne{} form, ensuring the correct number of
$d(d-2)-1$ propagating degrees of freedom. 

The derived \nPlusOne{} form of the bimetric field equations resembles
two copies of General relativity (GR). For comparison, the 3+1 split
is demonstrated for the bimetric spacetimes where the two sectors
share a common spherically symmetry. The \nPlusOne{} equations can
be used as a starting point for further analysis, for instance, in
numerical bimetric relativity.\vspace{-1mm}

\paragraph*{Structure of the paper.}

In the rest of this section we review some technical properties of
the field equations in GR and the HR bimetric theory. In section~\ref{sec:geom-mean},
we parametrize the metrics using the geometric mean, and decompose
the effective stress-energy tensors of the ghost-free bimetric potential.
After projecting the contracted Bianchi identities of the bimetric
effective stress-energy tensor, we obtain the \nPlusOne{} form of
the bimetric conservation law (the so-called secondary constraint
obtained using the Hamiltonian formalism \cite{Hassan:2011ea,Hassan:2018mbl}).
In section~\ref{sec:hr-N+1}, we give the complete set of the HR
field equations in standard \nPlusOne{} form, also summarized in
Box~1 on page~\pageref{box-1}. In the case of spherically symmetry,
the reduced \nPlusOne{} equations are studied in section~\ref{sec:ssym}.
The paper ends with a short summary and discussion. \vspace{1mm}

\noindent \emph{Notation}. The tilde indicates the most of the variables
in the $f$-sector. \ifColors As a visual aid, the variables of different
sectors are color-highlighted as in Figure~\ref{fig:gm-param}.\else
The preprint of this paper with the color-highlighted variables can
be found as an ancillary file on arXiv.\fi

\subsection{The \nPlusOne{} formalism in GR}

The Einstein-Hilbert action for a metric $\gMet$ is given as,
\begin{align}
\mathcal{S}_{\mathrm{EH}}[\gMet] & =\int\dd^{d}x\,\sqrt{-\gMet}\,\bigg[\frac{1}{2\gKappa}\,\gRicci+\mathcal{L}_{\gMet}^{\mathrm{m}}\bigg],\label{eq:EH-action}
\end{align}
where $\gRicci$ denotes the Ricci scalar for $\gMet$, and $\mathcal{L}_{\gMet}^{\mathrm{m}}$
denotes the Lagrangian density of matter fields minimally coupled
to gravity. The parameter $\gKappa$ is Einstein's gravitational constant.
Varying (\ref{eq:EH-action}) with respect to $\gMet$ gives the Einstein
field equations (here in operator form),
\begin{equation}
\tud{\gEinst}{\mu}{\nu}=\gKappa\tud{\gTse}{\mu}{\nu},\label{eq:GR-eom}
\end{equation}
where $\gEinst$ and $\gTse$ denote the Einstein tensor and the stress-energy
tensor, respectively,
\begin{equation}
\tud{\gEinst}{\mu}{\nu}\coloneqq\tud{\gRicci}{\mu}{\nu}-\frac{1}{2}\gRicci\delta_{\nu}^{\mu},\qquad\tud{\gTse}{\mu}{\nu}\coloneqq\frac{-2}{\sqrt{-\gMet}}\gMet^{\mu\rho}\frac{\partial(\sqrt{-\gMet}\thinspace\mathcal{L}_{\gMet}^{\mathrm{m}})}{\partial\gMet^{\rho\nu}}.\label{eq:def-T}
\end{equation}
Both $\tud{\gEinst}{\mu}{\nu}$ and $\tud{\gTse}{\mu}{\nu}$ are symmetric
(self-adjoint) with respect to $\gMet$. The Einstein tensor satisfies
the contracted Bianchi identity $\gCD_{\mu}\tud{\gEinst}{\mu}{\nu}=0$
which holds for any $\gMet$. Due to the contracted Bianchi identity,
the matter fields obey the conservation laws $\gCD_{\mu}\tud{\gTse}{\mu}{\nu}=0$.

The kinematical and dynamical parts of the metric field $\gMet$ can
be separated using the \nPlusOne{} decomposition; here we follow
the prescription from York \cite{York:1979aa}. One starts by foliating
the spacetime into a family of spacelike hypersurfaces $\{\Sigma\}$
with the future-pointing timelike unit normal $\vec{n}$ on $\{\Sigma\}$
satisfying $n_{\mu}n^{\mu}=-1$ with respect to $\gMet$. The geometrical
objects are then projected onto the slices using the operator,
\begin{equation}
\tud{\Proj}{\mu}{\nu}\coloneqq\tud{\delta}{\mu}{\nu}+n^{\mu}n_{\nu}.
\end{equation}
The metric induced on the spatial slices reads,
\begin{equation}
\gSp_{\mu\nu}\coloneqq\tud{\Proj}{\rho}{\mu}\tud{\Proj}{\sigma}{\nu}\gMet_{\rho\sigma}=\gMet_{\mu\nu}+n_{\mu}n_{\nu}.\label{eq:kd-7}
\end{equation}
Adapting a suitable chart $x^{\mu}=(t,x^{i})$ we can write $n=-\gLapse\dd t$,
$\vec{n}=\gLapse^{-1}(\partial_{t}-\gLapse^{i}\partial_{i})$, or,
\begin{equation}
n_{\mu}=\big(-\gLapse,0\big),\qquad n^{\mu}=\big(\gLapse^{-1},-\gLapse^{-1}\gShift{}^{i}\big),
\end{equation}
where $\gLapse$ and $\gShift^{i}$ are respectively the standard
lapse function and shift vector associated with the foliation. In
this chart, the metric decomposes according to (\ref{eq:kd-7}) as,
\begin{equation}
\gMet=-\gLapse^{2}\dd t^{2}+\gSp_{ij}\big(\dd x^{i}+\gShift{}^{i}\dd t\big)\big(\dd x^{j}+\gShift{}^{j}\dd t\big).\label{eq:kd-6}
\end{equation}
The lapse $\gLapse$ is a strictly positive scalar field (enforcing
$n^{0}>0$), the shift $\gShift{}^{i}$ is a purely spatial vector
field, and $\gSp_{ij}$ is a spatial Riemannian metric (the shift
vector $\gShift{}^{i}$ is confined to the spatial hypersurface, and
its indices are lowered using $\gSp_{ij}$). Also, the inverse of
$\gSp$ is obtained through $\gSp^{ik}\gSp_{kj}=\delta_{j}^{i}$.
In matrix notation, the metric (\ref{eq:kd-6}) reads,
\begin{equation}
\gMet_{\mu\nu}=\begin{pmatrix}-\gLapse^{2}+\gShift^{k}\gSp_{kl}\gShift^{l} & \, & \gShift^{k}\gSp_{kj}\\
\gSp_{il}\gShift^{l} &  & \gSp_{ij}
\end{pmatrix}\!.\label{eq:g-N+1}
\end{equation}

\paragraph*{Shear operator.}

Given a spatial vector $\xi$, one can define the shear operator,
\begin{equation}
\Xi[\xi]:=\begin{pmatrix}1 & \, & 0\\
\xi &  & \sI
\end{pmatrix}\!,\qquad\tud{\Xi[\xi]}{\mu}{\nu}=\begin{pmatrix}1 & \, & 0\\
\xi^{i} &  & \tud{\delta}ij
\end{pmatrix}\!.
\end{equation}
Note that the shear is unimodular, $\det\Xi[\xi]=1$, and it forms
an Abelian group isomorphic to $(\mathbb{R}^{d-1},+)$ since $\Xi[\xi_{1}]\Xi[\xi_{2}]=\Xi[\xi_{1}+\xi_{2}]$,
$\Xi[0]=I$, and $\Xi[\xi]^{-1}=\Xi[-\xi]$.

Taking the shift vector $\gShiftVec$ as a shear factor, the metric
$\gMet$ can be expressed,
\begin{equation}
\gMet=\Xi[\gShiftVec]^{\tr}\begin{pmatrix}-\gLapse^{2} & 0\\
0 & \gSp
\end{pmatrix}\Xi[\gShiftVec].\label{eq:g-Xi}
\end{equation}
The inverse of $\gMet$ then simply follows from the properties of
$\Xi$,
\begin{equation}
\quad\gMet^{-1}=\Xi[-\gShiftVec]\begin{pmatrix}-\gLapse^{-2} & 0\\
0 & \gSp^{-1}
\end{pmatrix}\Xi[-\gShiftVec]^{\tr},\quad\gMet^{\mu\nu}=\begin{pmatrix}-\gLapse^{-2} & \, & \gShift^{j}\gLapse^{-2}\\
\gShift^{i}\gLapse^{-2} &  & \gSp^{ij}-\gShift^{i}\gShift^{j}\gLapse^{-2}
\end{pmatrix}\!.\label{eq:g-Xi-inv}
\end{equation}

\paragraph*{Moving frames.}

Two types of spacetime frames are used in this work. One is a Cauchy
adapted frame \cite{Choquet:2008gr} based on a coframe $\{\theta^{\mu}\}$
with a dual $\{\vec{\theta}_{\mu}\}$, where $\theta^{\mu}(\vec{\theta}_{\nu})=\delta_{\nu}^{\mu}$
and,\bSe
\begin{alignat}{2}
\theta^{\mu} & \coloneqq\tud{\Xi[\gShiftVec]}{\mu}{\nu}\dd x^{\nu}, & \qquad\vec{\theta}_{\mu} & \coloneqq\tud{\Xi[-\gShiftVec]}{\nu}{\mu}\partial_{\nu},\label{eq:cauchy-1}\\
\ \ \theta^{0} & =\dd t,\quad\theta^{i}=\dd x^{i}+\gShift{}^{i}\dd t, & \qquad\vec{\theta}_{0} & =\partial_{t}-\gShift^{i}\partial_{i},\quad\vec{\theta}_{i}=\partial_{i},\\
\shortintertext{\text{such that,}}\gMet & =-\gLapse^{2}\theta^{0}\theta^{0}+\gSp_{ij}\theta^{i}\theta^{j}, & \qquad\gMet_{\mu\nu} & =-\gLapse^{2}\tud{\theta}0{\mu}\tud{\theta}0{\nu}+\gSp_{ij}\tud{\theta}i{\mu}\tud{\theta}j{\nu}.\label{eq:g-cauchy}
\end{alignat}
\eSe Note that a Cauchy adapted frame is seemingly a `coordinate
frame' where the spacetime indices are abused to label the basis vectors.
Recognizing $\tud{\theta}{\mu}{\nu}=\tud{\Xi[\gShiftVec]}{\mu}{\nu}$
from (\ref{eq:cauchy-1}), or equivalently $\theta=\Xi[\gShiftVec]$
in matrix notation, the expression (\ref{eq:g-cauchy}) is the same
as (\ref{eq:g-Xi}). 

The other type is an orthonormal frame or a vielbein, $\{E^{A}\}$,
with a dual $\{\vec{E}_{A}\}$ where $E^{A}(\vec{E}_{B})=\delta_{B}^{A}$
and,\bSe
\begin{alignat}{4}
\hspace{3em}E^{\hat{0}} & \coloneqq & \,\gLapse\theta^{0} & =\gLapse\dd t, & \vec{E}_{\hat{0}} & \coloneqq & \,\gLapse^{-1}\vec{\theta}_{0} & =\gLapse^{-1}(\partial_{t}-\gLapse^{i}\partial_{i}),\\
E^{a} & \coloneqq & \,\tud{\gE}ai\theta^{i} & =\tud{\gE}ai(\dd x^{i}+\gShift{}^{i}\dd t),\quad & \vec{E}_{a} & \coloneqq & \,\tud{\gE}ia\vec{\theta}_{i} & =\tud{\gE}ia\partial_{i},\\
\shortintertext{\text{such that,}}\gMet & =\mathrlap{-E^{\hat{0}}E^{\hat{0}}+\delta_{ab}E^{a}E^{b},} &  &  & \gMet_{\mu\nu} & =\mathrlap{\eta_{AB}\tud EA{\mu}\tud EB{\nu}.}
\end{alignat}
\eSe Here we used the beginning Latin letters to denote the indices
in the local Lorentz frame. Observe that $\tud{\gE}ia$ denotes $\gE^{-1}$
in matrix notation, and the vector field $\vec{E}_{\hat{0}}$ is the
timelike unit normal $\vec{n}$. The spatial metric is accordingly
decomposed as $\gSp_{ij}=\delta_{ab}\tud{\gE}ai\tud{\gE}bj$ (or $\gSp=\gE^{\tr}\sEta\gE$)
using a vielbein $\gE^{a}=\tud{\gE}ai\dd y^{i}$ tangential to $\Sigma$.
Subsequently, the metric (\ref{eq:g-Xi}) becomes,
\begin{equation}
\gMet=\Xi[\gShiftVec]^{\tr}\begin{pmatrix}\gLapse & 0\\
0 & \gE
\end{pmatrix}^{\tr}\!\begin{pmatrix}-1 & 0\\
0 & \sEta
\end{pmatrix}\begin{pmatrix}\gLapse & 0\\
0 & \gE
\end{pmatrix}\Xi[\gShiftVec],
\end{equation}
or $\gMet=E^{\tr}\eta E$ in terms of a vielbein $E$ in the lower
triangular form,
\begin{equation}
E=\begin{pmatrix}\gLapse & 0\\
0 & \gE
\end{pmatrix}\Xi[\gShiftVec]=\begin{pmatrix}\gLapse & 0\\
\gE\gShiftVec & \gE
\end{pmatrix}\!.
\end{equation}
Beware that an arbitrary vielbein can be triangularized by a local
Lorentz transformation if and only if the apparent lapse for the metric
is real in a given coordinate chart (see Lemma~\ref{lemma:vielb}
in appendix \ref{app:proof-prop-1}). For a single metric, a general
coordinate transformation can be used to ensure that the lapse is
real. This might not be possible for two arbitrary metrics.

\paragraph*{Projection meta-operators.}

Any symmetric tensor field $X_{\mu\nu}$ can be decomposed as,
\begin{equation}
X_{\mu\nu}=\prho[X]\,n_{\mu}n_{\nu}\,+\,n_{\mu}\pjota[X]_{\nu}+\pjota[X]_{\mu}n_{\nu}+\pJota[X]_{\mu\nu},\vspace{-3ex}\label{eq:proj-X}
\end{equation}
where,\vspace{-2ex}\bSe\label{eq:proj-ops}
\begin{alignat}{4}
\prho & [X] & \coloneqq &  & n^{\rho} & X_{\rho\sigma}n^{\sigma}, &  & \text{is the perpendicular projection},\label{eq:proj-op-rho}\\
\pjota & [X]_{\mu} & \coloneqq & \  & -\tud{\Proj}{\rho}{\mu} & X_{\rho\sigma}n^{\sigma}, &  & \text{is the mixed projection, and}\label{eq:proj-op-j}\\
\pJota & [X]_{\mu\nu} & \ \coloneqq &  & \tud{\Proj}{\rho}{\mu} & X_{\rho\sigma}\tud{\Proj}{\sigma}{\nu}, & \qquad & \text{is the full projection onto }\Sigma.\label{eq:proj-op-J}
\end{alignat}
\eSe The trace of $X$ can be expressed by,
\begin{equation}
\tud X{\mu}{\mu}=\gMet^{\mu\nu}X_{\mu\nu}=\gSp^{ij}\pJota[X]_{ij}-\prho[X]=\tud{\pJota[X]}ii-\prho[X].\label{eq:proj-X-trace}
\end{equation}
For the stress-energy tensor $T_{\mu\nu}$, the physical interpretation
of the projections is following: $\prho[T]$ is the energy density,
$\pjota[T]^{\mu}$ is the momentum density vector, $\pjota[T]_{\mu}$
is the energy flux covector, and $\pJota[T]_{\mu\nu}$ is the stress
tensor, all of them determined by the Eulerian observers whose worldlines
are orthogonal to the hypersurfaces $\{\Sigma\}$. (The unit timelike
vector $\vec{n}$ can be regarded as the velocity field of observers,
called the Eulerian or fiducial observers, which are instantaneously
at rest in the slices.) The minus sign in the definition of $\pjota$
(\ref{eq:proj-op-j}) ensures that $\pjota[T]$ points in the direction
in which matter is flowing.

\paragraph*{Projection of the field equations.}

The operators (\ref{eq:proj-ops}) can be used to decompose the field
equations (\ref{eq:GR-eom}) with the help of the Gauss\textendash Codazzi\textendash Mainardi
relations.

Let $\gK_{ij}$ be the extrinsic curvature expressed by,
\begin{equation}
\gK_{ij}\coloneqq\usignK\mfrac 12\,\Lie_{\vec{n}}\gSp_{ij},\qquad\gK\coloneqq\gSp^{ij}\gK_{ij}.\label{eq:K-def}
\end{equation}
\ifShowSigns Here $\Lie$ denotes the Lie derivative and the sign
convention for $\gK_{ij}$ is indicated by a label. \else Here $\Lie_{\vec{n}}$
denotes the Lie derivative along the vector field $\vec{n}$. \fi
 The perpendicular and the mixed projection of the Einstein tensor
reads,\bSe\label{eq:rhoj-G}
\begin{alignat}{2}
\prho[\gEinst]_{\hphantom{i}} & =\,\, & \gLapse^{2}\gEinst^{00} & =\mfrac 12\big(\gR+\gK^{2}-\gK_{ij}\gK^{ij}\big),\\
\pjota[\gEinst]_{i} & =\,\, & -\gLapse\tud{\gEinst}0i & =\ifShowSigns\isignK\big(\gD_{j}\tud{\gK}ji-\gD_{i}\gK\big)\else\uisignK\gD_{j}\tud{\gK}ji\signK\gD_{i}\gK\fi,
\end{alignat}
\eSe where $\gR=\gSp^{ij}\gR_{ij}$ is the trace of the Ricci tensor
$\gR_{ij}$ defined by the spatial covariant derivatives $\gD_{i}$
compatible with $\gSp_{ij}$. Combining (\ref{eq:rhoj-G}) with the
respective projections of the stress-energy tensor yields the constraint
equations,\bSe\label{eq:GR-constr}
\begin{alignat}{3}
\gCC & \coloneqq\, & \prho[\gEinst-\gKappa\gTse]_{\hphantom{i}} & =\,\, & \mfrac 12\big(\gR+\gK^{2}-\gK_{ij}\gK^{ij}\big)-\gKappa\,\prho[\gTse] & \,=0,\label{eq:GR-scalar-C}\\
\gCC_{i} & \coloneqq\, & \uisignK\pjota[\gEinst-\gKappa\gTse]_{i} & =\, & \gD_{k}\gK^{k}{}_{i}-\gD_{i}\gK\signK\gKappa\,\pjota[\gTse]_{i}\ \  & \,=0.\label{eq:GR-vector-C}
\end{alignat}
\eSe Equation (\ref{eq:GR-scalar-C}) is called the scalar or Hamiltonian
constraint while (\ref{eq:GR-vector-C}) is called the vector or momentum
constraint; they must be satisfied on the initial hypersurface.

\newpage

To end up with a more robust form of the evolution equations, we decompose
the field equations (\ref{eq:GR-eom}) based on the Ricci tensor (such
a procedure is due to York \cite{York:1979aa}),
\begin{equation}
\tud{\gRicci}{\mu}{\nu}=\gKappa\bigg(\tud{\gTse}{\mu}{\nu}-\frac{1}{d-2}\tud{\gTse}{\sigma}{\sigma}\tud{\delta}{\mu}{\nu}\bigg).\label{eq:GR-eom-Ricci}
\end{equation}
The full spatial projection of the $d$-dimensional Ricci tensor gives,
\begin{equation}
\pJota[\gRicci]_{ij}=\isignK\Lie_{\vec{n}}\gK_{ij}-\gLapse^{-1}\gD_{i}\gD_{j}\gLapse+\gR_{ij}-2\gK_{ik}\tud{\gK}kj+\gK\gK_{ij}.\label{eq:proj-R}
\end{equation}
Combining (\ref{eq:proj-R}) with the projection of the stress-energy
tensor, then expanding the Lie derivatives in terms of $\vec{n}=\gLapse^{-1}\left(\partial_{t}-\gShift{}^{i}\partial_{i}\right)$,
we get the evolution equations,\bSe\label{eq:GR-evol}
\begin{align}
\partial_{t}\gSp_{ij} & =\Lie_{\gShiftVec}\gSp_{ij}\signK2\gLapse\gK_{ij},\label{eq:GR-evol-1}\\
\partial_{t}\gK_{ij} & =\Lie_{\gShiftVec}\gK_{ij}\signK\gD_{i}\gD_{j}\gLapse\isignK\gLapse\,\big[\gR_{ij}-2\gK_{ik}\gK^{k}{}_{j}+\gK\gK_{ij}\big]\nonumber \\
 & \qquad\signK\gLapse\,\gKappa\Big\{\,\pJota[\gTse]_{ij}-\frac{1}{d-2}\Big(\tud{\pJota[\gTse]}kk-\prho[\gTse]\Big)\gSp_{ij}\,\Big\},\label{eq:GR-evol-2}
\end{align}
\eSe where the trace of $\gTse$ was decomposed using (\ref{eq:proj-X-trace}).
The equations (\ref{eq:GR-evol}) will be hereinafter referred to
be in \emph{standard} \nPlusOne{} \emph{form}.

\subsection{Bimetric field equations}

\label{sec:bim-fe}

We first consider a general class of bimetric actions consisting of
two Einstein-Hilbert terms $\mathcal{S}_{\mathrm{EH}}[\gMet]$ and
$\mathcal{S}_{\mathrm{EH}}[\fMet]$ each coupled to separate matter
fields, and the interaction term $\mathcal{L}_{\gMet,\fMet}^{\mathrm{int}}$
that depends on the scalar invariants of the operator $\gMet^{\mu\rho}\fMet_{\rho\nu}$
(written $\gMet^{-1}\fMet$ in matrix notation).\footnote{In the interacting case, the full diffeomorphism symmetry of two independent
Einstein-Hilbert terms is necessarily reduced to a diagonal subgroup
which restricts the potential to be dependent on $\gMet^{-1}\fMet$
\cite{Damour:2002ws}.} The bimetric action of such a class reads (in an arbitrary dimension
$d$),
\begin{align}
\mathcal{S} & =\int\dd^{d}x\sqrt{-\gMet}\,\bigg[\frac{1}{2\gKappa}\,\gRicci+\mathcal{L}_{\gMet}^{\mathrm{m}}\bigg]+\int\dd^{d}x\sqrt{-\fMet}\,\bigg[\frac{1}{2\fKappa}\,\fRicci+\mathcal{L}_{\fMet}^{\mathrm{m}}\bigg]\nonumber \\
 & \qquad\qquad+\int\dd^{d}x\sqrt{-\gMet}\,\mathcal{L}_{\gMet,\fMet}^{\mathrm{int}}(\gMet^{-1}\fMet),\label{eq:bim-action}
\end{align}
where $\gKappa$ and $\fKappa$ are the gravitational constants for
two sectors. By varying the action with respect to $\gMet$ and $\fMet$,
one obtains the field equations,\bSe\label{eq:bim-eom}
\begin{align}
\tud{\gEinst}{\mu}{\nu} & =\gKappa\tud{\gVse}{\mu}{\nu}+\gKappa\tud{\gTse}{\mu}{\nu},\\[1ex]
\tud{\fEinst}{\mu}{\nu} & =\fKappa\tud{\fVse}{\mu}{\nu}+\fKappa\tud{\fTse}{\mu}{\nu},
\end{align}
\eSe where $\gTse$ and $\fTse$ are the stress-energy tensors of
the matter fields, while $\gVse$ and $\fVse$ are the effective stress-energy
tensors of the bimetric potential $\mathcal{L}_{\gMet,\fMet}^{\mathrm{int}}(\gMet^{-1}\fMet)$,\bSe\label{eq:TV-defs}
\begin{alignat}{2}
\tud{\gTse}{\mu}{\nu} & \coloneqq\frac{-2}{\sqrt{-\gMet}}\gMet^{\mu\rho}\frac{\partial\big(\sqrt{-\gMet}\,\mathcal{L}_{\gMet}^{\mathrm{m}}\big)}{\partial\gMet^{\rho\nu}}, & \qquad\tud{\gVse}{\mu}{\nu} & \coloneqq\frac{-2}{\sqrt{-\gMet}}\gMet^{\mu\rho}\frac{\partial\big[\sqrt{-\gMet}\,\mathcal{L}_{\gMet,f}^{\mathrm{int}}(\gMet^{-1}\fMet)\big]}{\partial\gMet^{\rho\nu}},\\[1ex]
\tud{\fTse}{\mu}{\nu} & \coloneqq\frac{-2}{\sqrt{-\fMet}}\fMet^{\mu\rho}\frac{\partial\big(\sqrt{-\fMet}\,\mathcal{L}_{\fMet}^{\mathrm{m}}\big)}{\partial\fMet^{\rho\nu}}, & \tud{\fVse}{\mu}{\nu} & \coloneqq\frac{-2}{\sqrt{-\fMet}}\fMet^{\mu\rho}\frac{\partial\big[\sqrt{-\gMet}\,\mathcal{L}_{\gMet,f}^{\mathrm{int}}(\gMet^{-1}\fMet)\big]}{\partial\fMet^{\rho\nu}}.
\end{alignat}
\eSe\newpage 

\noindent The dependence of $\mathcal{L}_{\gMet,\fMet}^{\mathrm{int}}$
only on $\gMet^{-1}\fMet$, combined with the definitions of $\gVse$
and $\fVse$ from (\ref{eq:TV-defs}), implies the following algebraic
identities,\bSe\label{eq:bim-ids}
\begin{align}
\gMet^{\mu\rho}\frac{\partial\mathcal{L}_{\gMet,\fMet}^{\mathrm{int}}(\gMet^{-1}\fMet)}{\partial\gMet^{\rho\nu}}+\fMet^{\mu\rho}\frac{\partial\mathcal{L}_{\gMet,\fMet}^{\mathrm{int}}(\gMet^{-1}\fMet)}{\partial\fMet^{\rho\nu}} & =0,\label{eq:id-alg1}\\
\sqrt{-\gMet}\,\tud{\gVse}{\mu}{\nu}+\sqrt{-\fMet}\,\tud{\fVse}{\mu}{\nu}-\sqrt{-\gMet}\,V\,\tud{\delta}{\mu}{\nu} & =0,\label{eq:id-alg2}
\end{align}
and the differential identity,
\begin{equation}
\sqrt{-\gMet}\,\gCD_{\mu}\gVse{}^{\mu}{}_{\nu}+\sqrt{-\fMet}\,\fCD_{\mu}\fVse{}^{\mu}{}_{\nu}=0,\label{eq:id-damour}
\end{equation}
\eSe where $\gCD_{\mu}$ and $\fCD_{\mu}$ are the covariant derivatives
compatible with $\gMet$ and $\fMet$, respectively. The identity
(\ref{eq:id-alg2}) was first proved in \cite{Hassan:2014vja}. The
identity (\ref{eq:id-damour}) was shown in \cite{Damour:2002ws}
assuming that the action (\ref{eq:bim-action}) is invariant under
the diagonal subgroup of diffeomorphisms. In particular, it can be
proved that the differential identity (\ref{eq:id-damour}) is a consequence
of the definitions (\ref{eq:TV-defs}) provided that $\mathcal{L}_{\gMet,\fMet}^{\mathrm{int}}$
is a scalar function of $\gMet^{-1}\fMet$.

Assuming that $\gCD_{\mu}\tud{\gTse}{\mu}{\nu}=0$ and $\fCD_{\mu}\tud{\fTse}{\mu}{\nu}=0$,
the equations of motion (\ref{eq:bim-eom}) imply the Bianchi constraints,
\begin{equation}
\gCD_{\mu}\tud{\gVse}{\mu}{\nu}=0,\qquad\fCD_{\mu}\tud{\fVse}{\mu}{\nu}=0,\label{eq:bianchi}
\end{equation}
which are not independent according to (\ref{eq:id-damour}). The
equations (\ref{eq:bianchi}) will be hereinafter referred to as the
\emph{bimetric conservation law}.

\subsection{The ghost-free bimetric potential}

The absence of ghosts in the Hassan\textendash Rosen bimetric theory
is ensured by the interaction potential of the following form \cite{Hassan:2011zd},
\begin{equation}
\mathcal{L}_{\gMet,\fMet}^{\mathrm{int}}(\gMet^{-1}\fMet)=V(S)\coloneqq\signV\betaSum\,e_{n}(S),\label{eq:V}
\end{equation}
where $S$ is the principal square root of $\gMet^{-1}\fMet$, denoted
as $S=(\gMet^{-1}\fMet)^{1/2}$. In particular, the principal branch
of the square root provides an unambiguous definition of the HR theory
that guaranties the existence of a spacetime interpretation \cite{Hassan:2017ugh}. 

The negative sign convention for the interaction term places the
bimetric stress-energy contributions $\gVse$ and $\fVse$ on the
gravitational (left) side of the field equations, while the positive
sign places $\gVse$ and $\fVse$ on the matter (right) side.

The bimetric potential is parametrized by a set of real constants
$\{\beta_{n}\}$. The scale is given by $\betaScale$, making the
$\beta$-parameters dimensionless (in the geometrized unit system).
The coefficients $e_{n}(S)$ in (\ref{eq:V}) are the elementary symmetric
polynomials, which are the scalar invariants of $S$ obtained through
the generating function \cite{macdonald:1998a},
\begin{equation}
E(t,X)=\det(I+tX)=\sum_{n=0}^{\infty}e_{n}(X)\,t^{n}.\label{eq:e-genf}
\end{equation}
Note that $e_{n}(X)=0$ for all $n$ above the rank of $X$ due to
the Cayley-Hamilton theorem. This implies that the summation in (\ref{eq:V})
stops at $n=d$ for the nonsingular $S$.

\paragraph*{Elementary symmetric polynomials.}

\global\long\def\dimD{D}
 We summarize some important properties of the elementary symmetric
polynomials that are used throughout this work. Let $\tud X{\ixA}{\ixB}$
be an invertible operator on a vector space of an arbitrary dimension
$\dimD$. From (\ref{eq:e-genf}) we have,
\begin{equation}
e_{n}(X)=e_{\dimD}(X)e_{\dimD-n}(X^{-1}),\quad e_{\dimD}(X)e_{n}(X^{-1})=e_{\dimD-n}(X),\quad e_{\dimD}(X)=\det X.\label{eq:dual}
\end{equation}
A particularly useful expression for the elementary symmetric polynomials
reads,
\begin{equation}
e_{n}(X)=\tud X{[\ixA_{1}}{\ixA_{1}}\tud X{\ixA_{2}}{\ixA_{2}}\cdots\tud X{\ixA_{n}]}{\ixA_{n}}.\label{eq:e-def2}
\end{equation}
Consider now $\tud A{[\ixA_{1}}{\ixA_{1}}\tud B{\ixA_{2}}{\ixA_{2}}\cdots\tud B{\ixA_{n}]}{\ixA_{n}}$.
The expansion of the antisymmetrizer gives,
\begin{gather}
\tud A{[\ixA_{1}}{\ixA_{1}}\tud B{\ixA_{2}}{\ixA_{2}}\cdots\tud B{\ixA_{n}]}{\ixA_{n}}=\frac{1}{n}\Tr\left[A\,Y_{n-1}(B)\right],\label{eq:esp-1}\\
\shortintertext{\text{where we introduced,}\vspace{-1ex}}Y_{n}(X)\coloneqq\sum_{k=0}^{n}(-1)^{n+k}e_{k}(X)\,X^{n-k}.\label{eq:Yn-def}
\end{gather}
The function $Y_{n}$ satisfies the recursive relation (which can
be used as a definition of $Y_{n}$),
\begin{equation}
\,Y_{n}(X)=e_{n}(X)\,I-X\,Y_{n-1}(X),\quad Y_{0}(X)=I,\quad Y_{-1}(X)=0\,.\label{eq:YN-rec}
\end{equation}
The Cayley-Hamilton theorem can be written $Y_{\dimD}(X)=0$. From
(\ref{eq:YN-rec}) and (\ref{eq:dual}) follows,
\begin{equation}
Y_{n}(X)=e_{n}(X)\,I-e_{\dimD}(X)\,Y_{\dimD-n}(X^{-1}).\label{eq:id-alg2-alt}
\end{equation}
Using (\ref{eq:esp-1}) one obtains the following identities involving
a vector $u$ and covector $\omega$ (these relations occur in the
decompositions of Lorentz transformations and square roots),\bSe\label{eq:en_split}
\begin{align}
e_{n}(X+u\,\omega) & =e_{n}(X)+\omega\,Y_{n-1}(X)\,u,\\
e_{n}\big(X(I+u\,\omega)\big) & =e_{n}(X)\,(1+\omega u)+\omega\,X\,Y_{n-1}(X)\,u.
\end{align}
\eSe Setting $A=B=X$ in (\ref{eq:esp-1}) gives Newton's identities,
\begin{align}
e_{n}(X) & =\frac{1}{n}\Tr\left[X\,Y_{n-1}(X)\right]=\frac{1}{n}\sum_{k=1}^{n}(-1)^{k-1}e_{n-k}(X)\,\Tr(X^{k}).\label{eq:Newton}
\end{align}
Moreover, the recursive relation (\ref{eq:YN-rec}) implies the following
identities for traces, which are Newton's identities in disguise,
\begin{alignat}{2}
\Tr Y_{n}(X) & =(\dimD-n)\,e_{n}(X), & \qquad\Tr Y_{\dimD-n}(X^{-1}) & =n\,e_{n}(X)\,\det X.\label{eq:dual-ids}
\end{alignat}
Applying the derivative $\delta\Tr(X^{n})=n\Tr(X^{n-1}\delta X)$
on (\ref{eq:Newton}) gives,
\begin{equation}
\delta e_{n}(X)=\Tr\big[Y_{n-1}(X)\,\delta X\big]=\sum_{k=1}^{n}(-1)^{k-1}e_{n-k}(X)\,\Tr(X^{k-1}\delta X).\label{eq:var-en-X}
\end{equation}
Hence, $Y_{n}(X)$ can be equally defined as the derivative of $e_{n+1}(X)$,
\begin{equation}
Y_{n}(X)=\frac{\partial e_{n+1}(X)}{\partial X^{\tr}}.\label{eq:Yn-def2}
\end{equation}

\paragraph*{The HR field equations. }

The HR bimetric field equations have the form (\ref{eq:bim-eom})
with the following effective stress-energy contributions of the ghost-free
bimetric potential (\ref{eq:V}),\bSe\label{eq:Vgf}
\begin{align}
\gVse & =\usignV\betaSum\,Y_{n}(S),\\
\fVse & =\usignV\betaSum\,Y_{d-n}(S^{-1}).
\end{align}
\eSe The function $Y_{n}(S)$ is defined in (\ref{eq:Yn-def}), (\ref{eq:YN-rec}),
or (\ref{eq:Yn-def2}). The two contributions in (\ref{eq:Vgf}) are
not independent because they satisfy the identities (\ref{eq:id-alg2-alt})
and (\ref{eq:bim-ids}).

%% file: sec-20.tex
\section{Usage of the geometric mean}

\label{sec:geom-mean}

In this section we parametrize the metrics relative to the geometric
mean and obtain the projections of the bimetric stress-energy tensors
$\gVse$ and $\fVse$ as determined by the respective Eulerian observers.
For this we utilize two properties of the bimetric potential.

(i) The class of bimetric actions (\ref{eq:bim-action}) can exhibit
a duality in vacuum under the interchange $\gMet\leftrightarrow\fMet$
and $\gKappa\leftrightarrow\fKappa$. In the HR theory, the duality
is explicit under the additional exchange $\beta_{n}\leftrightarrow\beta_{d-n}$
as a consequence of (\ref{eq:dual}). The practical use is that we
need to derive only the relations for one sector and then simply apply
the results to the other, provided that the symmetry $\gMet\leftrightarrow\fMet$
is not algebraically broken (for example, when doing the \nPlusOne{}
decomposition). One way to achieve this is by using vielbeins.

(ii) The function space of the generating function (\ref{eq:e-genf})
is spanned by the polynomial basis $\{t^{n}\}$. In light of this,
the bimetric potential (\ref{eq:V}) and its stress-energy contributions
(\ref{eq:Vgf}) can be seen as a linear combination of the span $\{\beta_{n}$\},
where the elementary symmetric polynomials $e_{n}(S)$ and their derivatives
$Y_{n}(S)$ act as the coefficients. Consequently, the apparent basis
$\pm\betaSum$ can be neglected in the intermediate calculations;
namely, we can assume without loss of generality that,
\begin{equation}
\boxed{\,\ V=e_{n}(S),\qquad\gVse=Y_{n}(S),\qquad\fVse=Y_{d-n}(S^{-1}),\ }
\end{equation}
and recover $\betaSum$ (together with the chosen sign) at the end
when stating the complete set of equations. Note also that the summation
range is omitted from $\betaSum$ in (\ref{eq:V}) and (\ref{eq:Vgf})
since the vanishing $e_{n}(S)$ and $Y_{n}(S)$ will truncate the
sum by killing all unnecessary terms depending on the rank of $S$;
in particular, $e_{n>d}(S)=0$ and $Y_{n>d-1}(S)=0$. The summation
range will self-adapt to the lower dimension of the spatial hypersurface
upon the projection of the field equations using the \nPlusOne{}
decomposition, 

\subsection{Parametrization}

\label{ssec:param}

\begin{figure}
\noindent \begin{centering}
\includegraphics[scale=0.92]{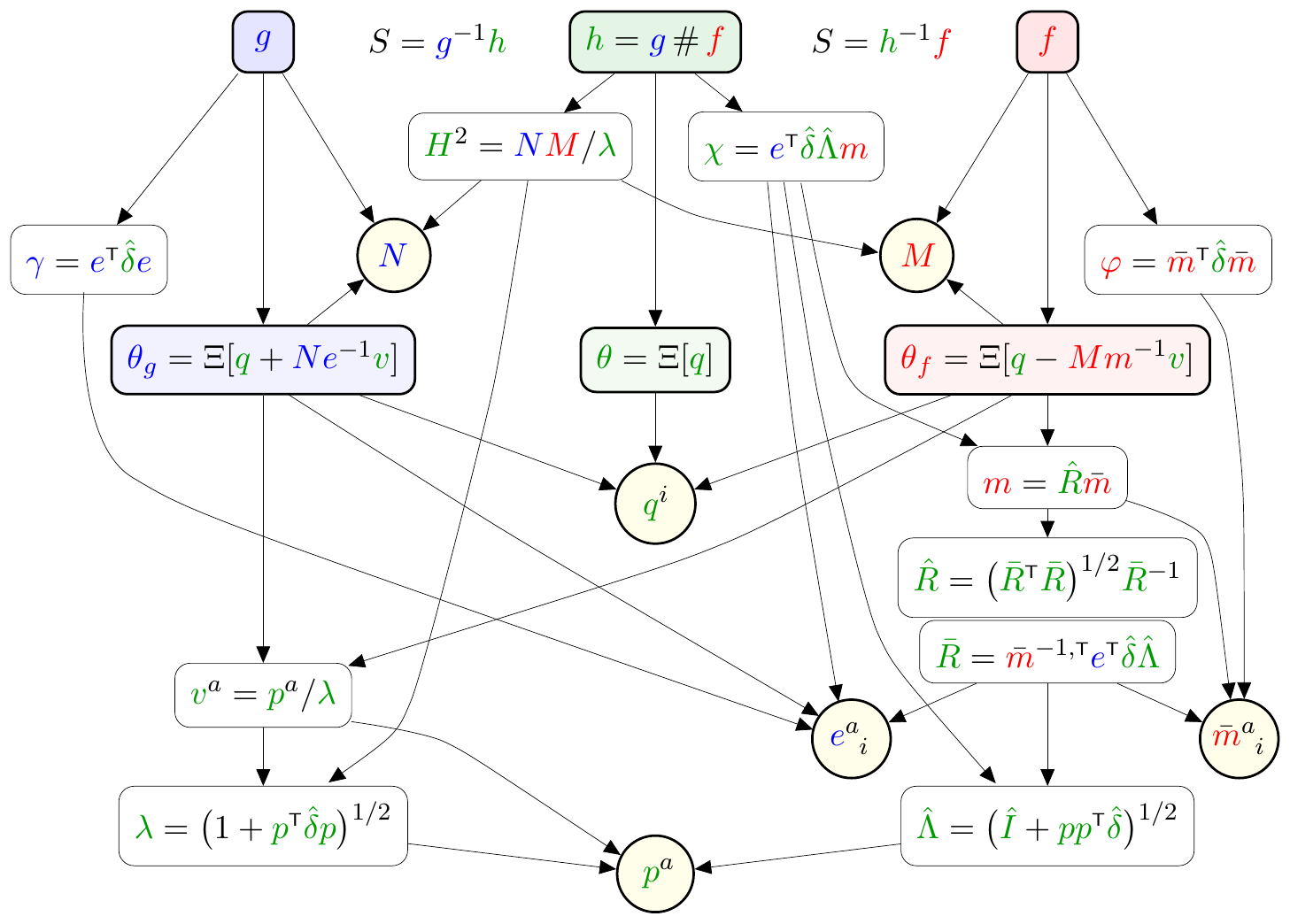}\vspace{-2ex}
\par\end{centering}
\caption{\label{fig:param-algo}The dependency graph for the parametrization
(see sec.~\ref{ssec:param} for explanations).}
\end{figure}

We work in a Cauchy frame $\theta=\Xi[\hShiftVec]$ adapted to $\hMet$
that is parametrized by a shift vector $\hShiftVec$. The shape of
the metric $\gMet$ is given by a lapse function $\gLapse$ and a
spatial vielbein $\gE$ that defines the spatial metric $\gSp=\gE^{\tr}\sEta\gE$.
The metric $\fMet$ is similarly parametrized by a lapse function
$\fLapse$ and a spatial vielbein $\bar{\fE}$ that defines the spatial
metric $\fSp=\bar{\fE}^{\tr}\sEta\bar{\fE}$. The lapses and the spatial
metrics are respectively perpendicular and tangential projections
on the spacelike hypersurfaces $\{\Sigma\}$. The mixed projections
are given by different shift vectors which define the respective Cauchy
frames. The overview of the parametrization as a dependency graph
is shown in Figure \ref{fig:param-algo} (which also gives the evaluation
order).\newpage

\noindent The Cauchy frames adapted to $\hMet$, $\gMet$ and $\fMet$
reads,
\begin{equation}
\hSector{\theta}\coloneqq\Xi\big[\hShift\big],\qquad\gSector{\theta}_{\gMet}\coloneqq\Xi\big[\hShift+\gLapse\gE^{-1}\sLp\sLtinv\big],\qquad\fSector{\theta}_{\fMet}\coloneqq\Xi\big[\hShift-\fLapse\fE^{-1}\sLp\sLtinv\big].\label{eq:Cauchy-frames}
\end{equation}
The separation between the frames is parametrized by a boost vector
$\sLp^{a}$ in the local Lorentz frame of $\hMet$, giving the Lorentz
factor,
\begin{equation}
\sLt\coloneqq\big(1+\sLp^{\tr}\sEta\sLp\big)^{1/2}.
\end{equation}
Moreover, given an arbitrary $\sLp$ and $\bar{\fE}$, the vielbein
$\fE$ in (\ref{eq:Cauchy-frames}) is obtained as,
\begin{equation}
\fE\coloneqq\sRs\bar{\fE},\quad\sRs=\big(\sRbar^{\tr}\sRbar\big)^{1/2}\sRbar^{-1},\quad\sRbar=\bar{\fE}^{-1,\tr}\gE^{\tr}\sEta\sLs,\label{eq:sim-fE}
\end{equation}
where $\sLs$ is the spatial part of the Lorentz boost given by,
\begin{equation}
\sLs\coloneqq\big(\sI+\sLp\sLp^{\tr}\sEta\big)^{1/2}.
\end{equation}
The relation (\ref{eq:sim-fE}) ensures that the spatial metric $\hSp$
of $\hMet$ is symmetric, 
\begin{equation}
\hSp=\gE^{\tr}\sEta\sLs\fE=\hSp^{\tr}.\label{eq:hSp}
\end{equation}
An important remark is that $\fSp=\fE^{\tr}\sEta\fE=\bar{\fE}^{\tr}\sEta\bar{\fE}$
since $\sRs$ is an orthogonal transformation $\sRs^{\tr}\sEta\sRs=\sEta$.
Note also that $\sLt$ is an eigenvalue of $\sLs$ with the eigenvector
$\sLv\coloneqq\sLp\sLtinv$, 
\begin{equation}
\sLs\sLv=\sLt\sLv=\sLp,\label{eq:Ls-ev}
\end{equation}
where $\sLv^{\tr}\sEta\sLv^{\tr}<1$. All other eigenvalues of $\sLs$
are 1, and $\det\sLs=\sLt$.\newpage

\noindent The resulting parametrization of the metrics reads,
\begin{equation}
\hMet=\hSector{\theta}^{\tr}\begin{pmatrix}-\hLapse^{2} & 0\\
0 & \hSp
\end{pmatrix}\hSector{\theta},\quad\ \gMet=\gSector{\theta}_{\gMet}^{\tr}\begin{pmatrix}-\gLapse^{2} & 0\\
0 & \gSp
\end{pmatrix}\gSector{\theta}_{\gMet},\quad\ \fMet=\fSector{\theta}_{\fMet}^{\tr}\begin{pmatrix}-\fLapse^{2} & 0\\
0 & \fSp
\end{pmatrix}\fSector{\theta}_{\fMet},
\end{equation}
where the lapse of $\hMet$ is given by,
\begin{equation}
\hLapse^{2}=\gLapse\fLapse\sLtinv.
\end{equation}
Notice that whenever the Cauchy frames coincide ($\sLt=1$, $\sLs=\sI$),
we have,
\begin{equation}
\hLapse=\gLapse\op{\#}\fLapse=\sqrt{\gLapse\fLapse},\quad\hSp=\gSp\op{\#}\fSp=\gSp(\gSp^{-1}\fSp)^{1/2},\quad\text{(for \ensuremath{\sLp=0})}.
\end{equation}

\begin{proposition}\label{prop:param}The parametrization based on
$\sLp$, $\hShiftVec$, $\gLapse$, $\fLapse$, $\gSp=\gE^{\tr}\sEta\gE$,
and $\fSp=\bar{\fE}^{\tr}\sEta\bar{\fE}$, yields all possible principal
square root realizations $S=(\gMet^{-1}\fMet)^{1/2}$ in the given
foliation.\end{proposition}

\noindent A detailed proof is given in appendix \ref{app:proof-prop-1}.
In brief, the effective shift vectors of $\gMet$ and $\fMet$ are,
\begin{equation}
\gShiftVec\coloneqq\hShift+\gLapse\gE^{-1}\sLp\sLtinv,\qquad\fShiftVec\coloneqq\hShift-\fLapse\fE^{-1}\sLp\sLtinv,
\end{equation}
yielding $\gShiftVec-\fShiftVec=\big(\gLapse\gE^{-1}+\fLapse\fE^{-1})\sLp\sLtinv$.
Subsequently, the parametrization is exhaustive based on the results
from section 3 in \cite{Hassan:2014gta}, and valid based on the theorem
from \cite{Hassan:2017ugh}.

The resulting square root $S=\gMet^{-1}\hMet=\hMet^{-1}\fMet$ reads,
\begin{equation}
S=\Xi[\hShiftVec]^{-1}\begin{pmatrix}\gLapse^{-1}\fLapse\sLtinv & \, & \sLt\gLapse^{-1}\sLv^{\tr}\sEta\fE\\
-\fLapse\sLtinv\gE^{-1}\sLv &  & \gE^{-1}\sLs^{-1}\fE
\end{pmatrix}\Xi[\hShiftVec].
\end{equation}
Notice how the algebraic form of $S$ is naturally given in the Cauchy
frame $\Xi[\hShiftVec]$ of $\hMet$. 

The square root can be further decomposed as,
\begin{equation}
S=Z^{-1}\begin{pmatrix}\sLtinv &  & \sLtinv\sLp^{\tr}\sEta\\
-\sLtinv\sLp & \, & \sLs-\sLtinv\sLp\sLp^{\tr}\sEta
\end{pmatrix}\begin{pmatrix}\fLapse\gLapse^{-1} & 0\\
0 & \fE\gE^{-1}
\end{pmatrix}Z,
\end{equation}
using the similarity transformation,\vspace{-1ex}
\begin{equation}
Z=\begin{pmatrix}\gLapse & 0\\
0 & \gE
\end{pmatrix}\begin{pmatrix}\sLtinv & 0\\
0 & \sI
\end{pmatrix}\begin{pmatrix}1 & 0\\
\hShiftVec & \sI
\end{pmatrix}.
\end{equation}
This form is useful when calculating matrix functions of $S$ since
$Z$ cancels out in traces and powers of $S$ (for instance, when
expanding the elementary symmetric polynomials $e_{n}$ or their derivatives
$Y_{n}$).

\paragraph*{Different views of the Lorentz frame.}

The boost vector $\sLv$ can be given with respect to $\gSp$ or $\fSp$
by changing the frame (basis) using the spatial vielbeins,
\begin{equation}
\sgn\coloneqq\gE^{-1}\sLv,\qquad\quad\sfn\coloneqq\fE^{-1}\sLv.\label{eq:sgn-def}
\end{equation}
We can equally recast $\sLs^{2}=\sI+\sLt^{2}\,\sLv\,\sLv^{\tr}\sEta$
by the corresponding similarity transformations,\bSe\label{eq:QQ-def}
\begin{alignat}{3}
\sgQ & \,\coloneqq\,\, & \gE^{-1} & \sLs^{2}\gE & \, & =\,\sI+\sLt^{2}\sgn\sgn^{\tr}\gSp,\\
\sfQ & \,\coloneqq\,\, & \fE^{-1} & \sLs^{2}\fE & \, & =\,\sI+\sLt^{2}\sfn\sfn^{\tr}\fSp.
\end{alignat}
\eSe From (\ref{eq:Ls-ev}), the eigenvectors of $\sgQ$ and $\sfQ$
are (\ref{eq:sgn-def}) having the eigenvalue $\sLt^{2}$,
\begin{equation}
\sgQ\sgn=\sLt^{2}\sgn,\qquad\quad\sfQ\sfn=\sLt^{2}\sfn,
\end{equation}
where $\sLt^{-2}=1-\sgn^{\tr}\gSp\sgn=1-\sfn^{\tr}\fSp\sfn$. 

To decompose the square roots of $\gSp^{-1}\fSp$ and $\fSp^{-1}\gSp$,
we define,\bSe\label{eq:DB-def}
\begin{alignat}{2}
\sgD & \coloneqq\fE^{-1}\sLs^{-1}\gE, & \qquad\quad\sgB & \coloneqq\sgD^{-1}=\gE^{-1}\sLs\fE=\sfD\sfQ,\\
\sfD & \coloneqq\gE^{-1}\sLs^{-1}\fE, & \qquad\quad\sfB & \coloneqq\sfD^{-1}=\fE^{-1}\sLs\gE=\sgD\sgQ,
\end{alignat}
\eSe so that $\gSp^{-1}\fSp$ and $\fSp^{-1}\gSp$ can be symmetrically
split into two components,\bSe
\begin{align}
\gSp^{-1}\fSp & =\sgD^{-1}\sfD=\sgB\sfD,\\
\fSp^{-1}\gSp & =\sfD^{-1}\sgD=\sfB\sgD.
\end{align}
\eSe Then, the spatial metric of $\hMet$ reads,
\begin{equation}
\hSp=\gSp\sgB=\gSp\sgD^{-1}=\fSp\sfB=\fSp\sfD^{-1}.
\end{equation}
Note that $\hSp$ is symmetric, $\hSp=\hSp^{\tr}$. Hence, any function
of $\sgD$ or $\sgB$ is symmetric with respect to $\gSp$, and any
function of $\sfD$ and $\sfB$ is symmetric with respect to $\fSp$. 

Finally, using (\ref{eq:DB-def}), the boost vectors $\sgn$ and $\sfn$
can be related across the sectors,\bSe
\begin{alignat}{4}
\sgn & =\gE^{-1}\sLv & \, & =\sLt\sfD\sfn, & \qquad\quad\sfn & =\fE^{-1}\sLv & \, & =\sLt\sgD\sgn,\\
\sfB\sgn & =\sgD\sgQ\sgn & \, & =\sLt\sfn, & \qquad\quad\sgB\sfn & =\sfD\sfQ\sfn & \, & =\sLt\sgn.
\end{alignat}
\eSe

\subsection{Decomposition of the potential}

The scalar density $\sqrt{-\gMet}\,V$ can be written,
\begin{align}
\gLapse\det\gE\;e_{n}(S) & =\gLapse\det\gE\;e_{n}(\gE^{-1}\sLs^{-1}\fE)+\fLapse\det\fE\;e_{d-n}(\fE^{-1}\sLs^{-1}\gE)\nonumber \\
 & =\gLapse\det\gE\;e_{n}(\sfD)+\fLapse\det\fE\;e_{d-n}(\sgD).\label{eq:NenS-2}
\end{align}
Here we used (\ref{eq:en_split}) together with the following relations
for $\sLs$,\bSe
\begin{alignat}{4}
\sLs^{2} & =\sI+\sLp\sLp^{\prime} &  & =\sI+\sLt^{2}\sLv\sLv^{\prime}, & \sLs^{-2} & =\sI-\sLt^{-2}\sLp\sLp^{\prime} &  & =\sI-\sLv\sLv^{\prime},\\
\sLs & =\sI+\frac{1}{\sLt+1}\sLp\sLp^{\prime} & \, & =\sI+\frac{\sLt^{2}}{\sLt+1}\sLv\sLv^{\prime}, & \qquad\sLs^{-1} & =\sI-\frac{1}{\sLt^{2}+\sLt}\sLp\sLp^{\prime} & \, & =\sI-\frac{\sLt}{\sLt+1}\sLv\sLv^{\prime},
\end{alignat}
\eSe where $\sLp^{\prime}\coloneqq\sLp^{\tr}\sEta$ and $\sLv^{\prime}\coloneqq\sLv^{\tr}\sEta$
denote the adjoint of of $\sLp$ and $\sLv$, respectively. 

Since $\sLtinv e_{n-1}(\sgB)=e_{d-n}(\sgD)\det(\fE\gE^{-1})$, we
can rewrite (\ref{eq:NenS-2}) as,
\begin{gather}
\gLapse V=\gLapse\sgV+\fLapse\sfV,\label{eq:NenS}\\
\shortintertext{\text{where,}}\sgV\coloneqq e_{n}(\sfD),\qquad\quad\sfV\coloneqq\sLtinv e_{n-1}(\sgB).\label{eq:sgV-def}
\end{gather}

\subsection{The bimetric stress-energy tensors}

Here we evaluate the projections of the bimetric stress-energy tensors
$\gVse$ and $\fVse$ as seen by their Eulerian observers. The projection
operators for the frames of $\hMet$, $\gMet$, and $\fMet$ are,
\begin{equation}
\hProj=\begin{pmatrix}0 &  & 0\\
\hShiftVec & \, & \sI
\end{pmatrix},\qquad\gProj=\hProj+\begin{pmatrix}0 & 0\\
\gLapse\gE^{-1}\sLv & 0
\end{pmatrix},\qquad\fProj=\hProj-\begin{pmatrix}0 & 0\\
\fLapse\fE^{-1}\sLv & 0
\end{pmatrix}.
\end{equation}
We first determine the projections of $\gVse=Y_{n}(S)$ relative to
the Eulerian observer of $\gMet$. The one-side projection of $S$
can be evaluated as,
\begin{equation}
\tud{\gProj}{\mu}{\sigma}\tud S{\sigma}{\nu}=\begin{pmatrix}0 & 0\\
0 & \tud{\sgB}ik
\end{pmatrix}\begin{pmatrix}1 & 0\\
\hShift^{k} & \tud{\delta}kj
\end{pmatrix},\qquad e_{n}(\gProj S)=e_{n}(\sgB).\label{eq:proj-g-S}
\end{equation}
Using (\ref{eq:proj-g-S}) together with the recursive relations (\ref{eq:YN-rec})
for $Y_{n}(S)$, we obtain,\bSe\label{eq:Vg-split}
\begin{alignat}{2}
\grho & =\prho[\gVse] & \, & =-e_{n}(\sgB),\\
\gjota_{i} & =\pjota[\gVse]_{i} & \, & =-\gSp_{ij}\tud{\sgQU}jk\sgn^{k},\\
\tud{\gJota}ik & =\pJota[\gVse\tud ]ik & \, & =\big\llbracket\,\sgV\sI-\sgQU+\gLapse^{-1}\fLapse\,\sgU\,\tud{\big\rrbracket}ik,
\end{alignat}
\eSe where the double brackets are used to enclose matrix expressions
when putting indices and,\bSe\label{eq:UQ-def}
\begin{equation}
\sgU\coloneqq\sLtinv Y_{n-1}(\sgB),\qquad\sfU\coloneqq\sfD\,Y_{n-1}(\sfD).
\end{equation}
The composite symbols for the mixed terms $\sgQU$ and $\sfQU$ are
defined as,
\begin{equation}
\sgQU\coloneqq\sgQ\sfU=\sgB\,Y_{n-1}(\sfD),\qquad\sfQU\coloneqq\sfQ\sgU=\sLtinv\sfQ\,Y_{n-1}(\sgB),
\end{equation}
\eSe where we used $\sgQ\sfD=\sfD\sfQ$ from (\ref{eq:QQ-def}) and
(\ref{eq:DB-def}). Observe that $\gSp_{ik}\sgU\tud{}kj$, $\gSp_{ik}\sgQU\tud{}kj$,
$\fSp_{ik}\sfU\tud{}kj$, and $\fSp_{ik}\sfQU\tud{}kj$ are symmetric
tensor fields.  The projections of $\fVse$ can be deduced from the
duality between the sectors by using (\ref{eq:dual}) and (\ref{eq:id-alg2-alt});
we obtain,\bSe\label{eq:Vf-split}
\begin{alignat}{2}
\frho & =\prho[\fVse] & \, & =-\sLt e_{n-1}(\sfD)\,\det(\gE\fE^{-1}),\\
\fjota_{i} & =\pjota[\fVse]_{i} & \, & =-\gjota_{i}\,\det(\gE\fE^{-1}),\\
\tud{\fJota}ik & =\pJota[\fVse\tud ]ik & \, & =\big\llbracket\,\sfV\sI-\sfQU+\fLapse^{-1}\gLapse\,\sfU\,\tud{\big\rrbracket}ik\,\det(\gE\fE^{-1}).
\end{alignat}
\eSe Having $\sgV=e_{n}(\sfD)$, it holds,
\begin{equation}
\grho+\sgV=\gjota_{k}\sgn^{k},\quad\text{or}\quad e_{n}(\sgB)-e_{n}(\sfD)=\sgn^{\tr}\gSp\sgQU\sgn,\label{eq:rho-v-identity}
\end{equation}
where $\sgU$ and $\sfU$ satisfy $\sgU\sgn=\sfU\sfn$ and $\sgn^{\tr}\gSp\sgQU\sgn=\sLt^{2}\sfn^{\tr}\fSp\sfU\sfn$.

We summarize the projections of $\gVse$ and $\fVse$, now given in
a slightly different form, \bSe
\begin{alignat}{2}
\grho & =-e_{n}(\sgB)=\sgn^{\tr}\gjota-\sgV, & \frho\,\frac{\det\fE}{\det\gE} & =-\sLt e_{n-1}(\sfD),\\
\gjota & =-\gSp\sgQU\sgn, & \fjota\,\frac{\det\fE}{\det\gE} & =-\gjota=\gSp\sgQU\sgn,\\
\gLapse\gJota & =\gLapse\,\big[\sgV\sI-\sgQU\big]+\fLapse\,\sgU, & \qquad\fLapse\fJota\,\frac{\det\fE}{\det\gE} & =\fLapse\,\big[\sfV\sI-\sfQU\big]+\gLapse\,\sfU.
\end{alignat}
\eSe Recall that the projections are implicitly spanned by $\pm\betaSum$.
For completeness, the components of $\gVse$ and $\fVse$ in terms
of (\ref{eq:sgV-def}) and (\ref{eq:UQ-def}) are given in appendix
\ref{app:VgVf_components}.

\subsection{The bimetric conservation law}

\label{subsec:bim-cons-law}

In GR, the conservation law $\gCD_{\mu}\tud{\gTse}{\mu}{\nu}=0$ needs
to be specifically imposed upon the \nPlusOne{} decomposition, otherwise
the constraint equations will fail to be preserved during the evolution
\cite{Gourgoulhon:2012trip}. For a class of bimetric theories having
the field equations (\ref{eq:bim-eom}), the effective stress-energy
tensor contributions (\ref{eq:TV-defs}) are included in the Bianchi
constraints (\ref{eq:bianchi}). Therefore, after putting the bimetric
field equations into the \nPlusOne{} form, we also need to assume
that the bimetric conservation law $\gCD_{\mu}\tud{\gVse}{\mu}{\nu}=0$
is specifically satisfied. 

In the following, we derive the projection of the conservation law
for the ghost-free bimetric potential $V=e_{n}(S)$. Let $\gVse$
be given in terms of the energy density $\grho$, the energy flux
covector $\gjota_{i}$, and the stress tensor $\tud{\gJota}ij$, all
of them measured as (\ref{eq:Vg-split}) by the Eulerian observer
relative to $\gMet$. A straightforward manipulation yields the following
projections of $\gCD_{\mu}\tud{\gVse}{\mu}{\nu}=0$ (summarized in
appendix \ref{app:proj-nabla-T}),\bSe\label{eq:proj-sv}
\begin{align}
\partial_{t}\grho & =\Lie_{\gShiftVec}\grho-\gLapse\gD_{i}\gjota^{i}-2\gjota^{i}\gD_{i}\gLapse\isignK\gLapse\gK\grho\isignK\gLapse\tud{\gK}ji\tud{\gJota}ij,\\
\partial_{t}\gjota_{i} & =\Lie_{\gShiftVec}\gjota_{i}-\gD_{j}\big[\gLapse\tud{\gJota}ji\big]-\grho\gD_{i}\gLapse\isignK\gLapse\gK\gjota_{i},
\end{align}
\eSe where the perpendicular projection of $\gCD_{\mu}\tud{\gVse}{\mu}{\nu}$
with respect to $\hMet$ can be identified as,
\begin{equation}
n^{\nu}\gCD_{\mu}\tud{\gVse}{\mu}{\nu}+\sgn^{i}\tud{\gProj}{\nu}i\gCD_{\mu}\tud{\gVse}{\mu}{\nu}=\partial_{t}\grho-\sgn^{i}\partial_{t}\gjota_{i}.
\end{equation}

\begin{lemma}\label{lemma:vgf-flux}The energy density and the energy
flux of the bimetric potential satisfy,\bSe\label{eq:lem1a}
\begin{align}
\partial_{t}\grho-\sgn^{i}\partial_{t}\gjota_{i} & =\frac{1}{2}\sgQU^{ij}\partial_{t}\gSp_{ij}-\frac{1}{2}\sfU^{ij}\partial_{t}\fSp_{ij},\label{eq:lem1a-1}\\
\partial_{k}\grho-\sgn^{i}\partial_{k}\gjota_{i} & =\frac{1}{2}\sgQU^{ij}\partial_{k}\gSp_{ij}-\frac{1}{2}\sfU^{ij}\partial_{k}\fSp_{ij}.\label{eq:lem1a-2}
\end{align}
\eSe Moreover, for the Lie flow along an arbitrary vector field $\qvf^{i}$,
it holds,\bSe\label{eq:lem1b}
\begin{align}
\Lie_{\qvf}\grho-\sgn^{i}\Lie_{\qvf}\gjota_{i} & =\hphantom{+}\tud{\sgQU}ij\gD_{i}\qvf^{j}-\tud{\sfU}ij\fD_{i}\qvf^{j},\label{eq:lem1b-1}\\
\Lie_{\qvf}\sgV-\gjota_{i}\Lie_{\qvf}\sgn^{i} & =-\tud{\sgQU}ij\gD_{i}\qvf^{j}+\tud{\sfU}ij\fD_{i}\qvf^{j}.\label{eq:lem1b-2}
\end{align}
\eSe\end{lemma}

\noindent The proof of the lemma is given in the first part of appendix
\ref{app:proj-nabla-Vg}. Useful choices for $\qvf^{i}$ are the shifts
$\gShift^{i}$ and $\fShift^{i}$, or the boost parameters $\sgn^{i}$
and $\sfn^{i}$, where $\Lie_{\sgn}\,\sgn=\Lie_{\sfn}\,\sfn=0$. 

Applying Lemma~\ref{lemma:vgf-flux} on (\ref{eq:proj-sv}) gives
the \nPlusOne{} form of the conservation law for $\gVse$ (the derivations
are in the second part of appendix \ref{app:proj-nabla-Vg}),
\begin{equation}
\tud{\sgU}ij\Big(\gD_{i}\sgn^{j}\signK\tud{\gK}ji\Big)+\tud{\sfU}ij\Big(\fD_{i}\sfn^{j}\isignK\tud{\fK}ji\Big)=\gD_{i}\big[\tud{\sgU}ij\sgn^{j}\big].\label{eq:bim-prop-constr-1}
\end{equation}
A more symmetric form that reflects the duality $\gMet\leftrightarrow\fMet$
is given by (\ref{eq:proj-constr-sym}), copied here,
\begin{align}
\frac{1}{2}\gD_{i}\big[\tud{\sgU}ij\sgn^{j}\big]+\frac{1}{2}\fD_{i}\big[\tud{\sfU}ij\sfn^{j}\big] & =\tud{\sgU}ij\bigg(\gD_{i}\sgn^{j}-\frac{1}{2}\frac{\partial_{i}\sqrt{\gSp}}{\sqrt{\gSp}}\sgn^{j}\signK\tud{\gK}ji\bigg)\nonumber \\
 & \qquad+\tud{\sfU}ij\bigg(\fD_{i}\sfn^{j}+\frac{1}{2}\frac{\partial_{i}\sqrt{\fSp}}{\sqrt{\fSp}}\sfn^{j}\isignK\tud{\fK}ji\bigg).\label{eq:proj-constr-sym-1}
\end{align}
Note that the covariant derivatives in (\ref{eq:proj-constr-sym-1})
can be easily converted to the partial derivatives of the respective
metric since $\gSp\sgU$ and $\fSp\sfU$ are symmetric.

%% file: sec-30.tex
\makeatletter
\long\def\@makecaption#1#2{%
  \vskip\abovecaptionskip
  \sbox\@tempboxa{#1. #2}%
  \ifdim \wd\@tempboxa >\hsize
    \small #1. #2\par
  \else
    \global \@minipagefalse
    \hb@xt@\hsize{\hfil\box\@tempboxa\hfil}%
  \fi
  \vskip\belowcaptionskip}
\makeatother

\section{The HR bimetric field equations in standard \nPlusOne{} form}

\label{sec:hr-N+1}

 The HR field equations (\ref{eq:bim-eom}) are two copies of (\ref{eq:GR-eom})
with the additional $\gVse$ and $\fVse$ that couple two sectors
through the identities (\ref{eq:bim-ids}). Therefore, to state the
HR field equations in the \nPlusOne{} form, we only need to include
the effective stress-energy contributions when projecting the particular
sector (using an appropriate timelike unit normal):\bSe
\begin{alignat}{3}
\grho & \coloneqq\prho[\gVse+\gTse],\qquad & \gjota & \coloneqq\pjota[\gVse+\gTse],\qquad & \gJota & \coloneqq\pJota[\gVse+\gTse],\\
\frho & \coloneqq\prho[\fVse+\fTse], & \fjota & \coloneqq\pjota[\fVse+\fTse], & \fJota & \coloneqq\pJota[\fVse+\fTse],
\end{alignat}
\eSe where $\gTse=0$ and $\fTse=0$ in the ``bimetric vacuum.''
The substitution is straightforward (it is mostly a ``copy \& paste''
of the GR results). The end result is the complete set of the HR equations
in standard \nPlusOne{} form, given below. 

The evolution equations for the dynamical pair $(\gSp_{ij},\gK_{ij})$
in the $\gMet$ sector reads,\bSe\label{eq:gf-evol}
\begin{align}
\partial_{t}\gSp_{ij} & =\Lie_{\hShift}\gSp_{ij}+\Lie_{\gLapse\sgn}\gSp_{ij}\signK2\gLapse\gK_{ij},\\[1ex]
\partial_{t}\gK_{ij} & =\Lie_{\hShift}\gK_{ij}+\Lie_{\gLapse\sgn}\gK_{ij}\signK\gD_{i}\gD_{j}\gLapse\isignK\gLapse\,\big[\gR_{ij}-2\gK_{ik}\gK^{k}{}_{j}+\gK\gK_{ij}\big]\nonumber \\
 & \qquad\quad\signK\gLapse\gKappa\Big\{\,\gSp_{ik}\tud{\gJota}kj-\frac{1}{d-2}\gSp_{ij}(\gJota-\grho)\,\Big\},
\end{align}
while the evolution equations for the dynamical pair $(\fSp_{ij},\fK_{ij})$
in the $\fMet$ sector reads,
\begin{align}
\partial_{t}\fSp_{ij} & =\Lie_{\hShift}\fSp_{ij}-\Lie_{\fLapse\sfn}\fSp_{ij}\signK2\fLapse\fK_{ij},\\[1ex]
\partial_{t}\fK_{ij} & =\Lie_{\hShift}\fK_{ij}-\Lie_{\fLapse\sfn}\fK_{ij}\signK\fD_{i}\fD_{j}\fLapse\isignK\fLapse\,\big[\fR_{ij}-2\fK_{ik}\fK^{k}{}_{j}+\fK\fK_{ij}\big]\nonumber \\
 & \qquad\quad\signK\fLapse\fKappa\Big\{\,\fSp_{ik}\tud{\fJota}kj-\frac{1}{d-2}\fSp_{ij}(\fJota-\frho)\,\Big\}.
\end{align}
\eSe The full spatial projections of $\fVse+\fTse$ and $\gVse+\gTse$
are given by,\bSe\label{eq:J-projs}
\begin{align}
\tud{\gJota}kj & =\big\llbracket\,\sgV\sI-\sgQU+\fLapse\gLapse^{-1}\sgU\,\tud{\big\rrbracket}kj+\pJota[\gTse\tud ]kj.\\[1ex]
\tud{\fJota}kj & =\frac{\sqrt{\gSp}}{\sqrt{\fSp}}\big\llbracket\,\sfV\sI-\sfQU+\gLapse\fLapse^{-1}\sfU\,\tud{\big\rrbracket}kj+\pJota[\fTse\tud ]kj.
\end{align}
The traces of $\gVse+\gTse$ and $\fVse+\fTse$ can be obtained from
(\ref{eq:dual-ids}),
\begin{align}
\gJota-\grho & =\usignV\betaSumL(d-n)\,e_{n}(S)+\pJota[\gTse\tud ]kk-\prho[\gTse],\\[0.5ex]
\fJota-\frho & =\usignV\frac{\sqrt{\gSp}}{\sqrt{\fSp}}\betaSumL n\,e_{n}(S)+\pJota[\fTse\tud ]kk-\prho[\fTse],
\end{align}
\eSe where from (\ref{eq:NenS-2}) and (\ref{eq:NenS}),
\begin{equation}
e_{n}(S)=e_{n}(\sfD)+\frac{\fLapse\det\fE}{\gLapse\det\gE}e_{d-n}(\sgD)=e_{n}(\sfD)+\frac{\fLapse}{\gLapse\sLt}e_{n-1}(\sgB).
\end{equation}
\newpage\noindent The constraint equations for the two sectors are
given by,\bSe\label{eq:g-cc}\bgroup
\setlength{\belowdisplayskip}{0pt}
\begin{alignat}{3}
2\gCC & \coloneqq\gR+\gK^{2}-\gK_{ij}\gK^{ij} & \, & \,- & \,2\gKappa\,\grho & =0,\label{eq:g-cc-scalar}\\
2\fCC & \coloneqq\fR+\fK^{2}-\fK_{ij}\fK^{ij} & \, & \,- & \,2\fKappa\,\frho & =0,\label{eq:f-cc-scalar}\\
\gCC_{i} & \coloneqq\gD_{k}\gK^{k}{}_{i}-\gD_{i}\gK & \, & \signK & \gKappa\,\gjota_{i} & =0,\label{eq:g-cc-vector}\\
\fCC_{i} & \coloneqq\fD_{k}\fK^{k}{}_{i}-\fD_{i}\fK & \, & \signK & \fKappa\,\fjota_{i} & =0,\label{eq:f-cc-vector}
\end{alignat}
\egroup\eSe where,\bSe\bgroup
\setlength{\abovedisplayskip}{0pt}
\begin{alignat}{2}
\grho & =\uisignV\betaSumL\,e_{n}(\sgB)+\prho[\gTse]\,, & \gjota_{i} & =\pjota[\gVse]_{i}+\pjota[\gTse]_{i}\,,\\
\frho & =\uisignV\frac{\sqrt{\gSp}}{\sqrt{\fSp}}\betaSumL\sLt e_{n-1}(\sfD)+\prho[\fTse]\,,\qquad & \fjota_{i} & =\pjota[\fVse]_{i}+\pjota[\fTse]_{i}\,,\\
0 & =\sqrt{\gSp}\,\pjota[\gVse]_{i}+\sqrt{\fSp}\,\pjota[\fVse]_{i}\,, & \pjota[\gVse]_{i} & =-\gSp_{ij}\tud{\sgQU}jk\sgn^{k}\,.\label{eq:f-cc-j-def}
\end{alignat}
\egroup\eSe The constraints for the mixed projections (\ref{eq:g-cc-vector})\textendash (\ref{eq:f-cc-vector})
are coupled by (\ref{eq:f-cc-j-def}) yielding,
\begin{align}
\hCC_{i} & \coloneqq\sqrt{\gSp}\Big\{\gKappainv\big(\gD_{k}\gK^{k}{}_{i}-\gD_{i}\gK\big)\signK\pjota[\gTse]_{i}\Big\}\nonumber \\
 & \qquad+\,\sqrt{\fSp}\Big\{\fKappainv\big(\fD_{k}\fK^{k}{}_{i}-\fD_{i}\fK\big)\signK\pjota[\fTse]_{i}\Big\}=0.\label{eq:h-cc-mixed}
\end{align}
The last equation can be used as a replacement for either (\ref{eq:g-cc-vector})
or (\ref{eq:f-cc-vector}), where the other becomes the equation of
motion for $\sLp$ (i.e., for $\sgn$ and $\sfn$). Observe that the
lapses $\gLapse$, $\fLapse$, and the shift vector $\hShiftVec$
of $\hMet$ are absent from the constraints, as expected. 

In addition to (\ref{eq:gf-evol})\textendash (\ref{eq:h-cc-mixed}),
we also have to assume specifically that (i) the conservation laws
$\gCD_{\mu}\tud{\gTse}{\mu}{\nu}=0$ and $\fCD_{\mu}\tud{\fTse}{\mu}{\nu}=0$
hold, and (ii) the conservation law for the bimetric potential $\gCD_{\mu}\tud{\gVse}{\mu}{\nu}=0$
holds. Failing to do so, the constraints will not be preserved during
the dynamical evolution by (\ref{eq:gf-evol}). Note that we do not
need to regard $\fCD_{\mu}\tud{\fVse}{\mu}{\nu}=0$ because of the
differential identity (\ref{eq:id-damour}). Consequently, the propagation
of the constraints requires that the equation (\ref{eq:bim-prop-constr-1})
is satisfied,
\begin{equation}
\tud{\sgU}ij\Big(\gD_{i}\sgn^{j}\signK\tud{\gK}ji\Big)+\tud{\sfU}ij\Big(\fD_{i}\sfn^{j}\isignK\tud{\fK}ji\Big)-\gD_{i}\big[\tud{\sgU}ij\sgn^{j}\big]=0.\label{eq:bim-prop-constr}
\end{equation}
The equation (\ref{eq:bim-prop-constr}) is equivalent to (and algebraically
the same as) the so-called \emph{secondary constraint} which is obtained
using the Hamiltonian formalism \cite{Hassan:2011ea,Hassan:2018mbl}.
In the context of the \nPlusOne{} decomposition employed here, it
is rather seen as the conservation law for the ghost-free bimetric
potential.

This concludes the HR bimetric field equations in standard \nPlusOne{}
form. An overview of the used variables is given in Table~\ref{tab:vars}.
A handy overview of the \nPlusOne{} equations is given in Box~1.
The counting of the degrees of freedom is given in Table~\ref{tab:dof}.
The counting begins with $2d(d-1)$ equations for the dynamical fields
$(\gSp_{ij},\gK_{ij})$ and $(\fSp_{ij},\fK_{ij})$. Besides, there
are also $d+1$ constraint equations, the bimetric conservation law
equation, and an overall gauge freedom which allows us to remove additionally
$d$ of the dynamical fields. As a result, we are left with $d(d-2)-1$
truly dynamical conjugate pairs, which is the number of the propagating
degrees of freedom in the HR theory.

\begin{table}
\caption{\label{tab:vars}An overview of the used \nPlusOne{} variables.}

\vspace{0.2ex}

\bgroup\centering\renewcommand{\arraystretch}{1.4}
\noindent \begin{centering}
\begin{tabular}{|r|l||r|l||l|}
\multicolumn{2}{c}{$\gSector{g\text{-sector}}$} & \multicolumn{2}{c}{$\fSector{f\text{-sector}}$} & \multicolumn{1}{l}{\emph{Comments}}\tabularnewline
\hline 
$\gSp\coloneqq$ & $\gE^{{\scriptscriptstyle \tr}}\sEta\gE$ & $\fSp\coloneqq$ & $\fE^{{\scriptscriptstyle \tr}}\sEta\fE$ & Spatial metric\tabularnewline
\hline 
$\sgn\coloneqq$ & $\gE^{-1}\sLv$ & $\sfn\coloneqq$ & $\fE^{-1}\sLv$ & Boost parameter\tabularnewline
\hline 
$\sgQ\coloneqq$ & $\gE^{-1}\sLs^{2}\gE$ & $\sfQ\coloneqq$ & $\fE^{-1}\sLs^{2}\fE$ & Spatial part of LLT\tabularnewline
\hline 
$\sgD\coloneqq$ & $\fE^{-1}\sLs^{-1}\gE$ & $\sfD\coloneqq$ & $\gE^{-1}\sLs^{-1}\fE$ & 1st half of $\gSp^{-1}\fSp$\tabularnewline
\hline 
$\sgB\coloneqq$ & $\sgD^{-1}=\gE^{-1}\sLs\fE$ & $\sfB\coloneqq$ & $\sfD^{-1}=\fE^{-1}\sLs\gE$ & 2nd half of $\gSp^{-1}\fSp$\tabularnewline
\hline 
 & $\sgB=\sfD\sfQ=\sgQ\sfD$ &  & $\sfB=\sgD\sgQ=\sfQ\sgD$ & Mixed terms\tabularnewline
\hline 
\multicolumn{1}{r}{} & \multicolumn{1}{l}{} & \multicolumn{1}{r}{} & \multicolumn{1}{l|}{} & \tabularnewline[-4ex]
\hline 
$\sgV\coloneqq$ & $\diamond\,e_{n}(\sfD)$ & $\sfV\coloneqq$ & $\diamond\,\sLtinv e_{n-1}(\sgB)$ & Potential$^{\ast}$\tabularnewline
\hline 
$\sgU\coloneqq$ & $\diamond\,\sLtinv Y_{n-1}(\sgB)$ & $\sfU\coloneqq$ & $\diamond\,\sfD\,Y_{n-1}(\sfD)$ & Derivatives$^{\ast}$\tabularnewline
\hline 
$\sgQU\coloneqq$ & $\diamond\,\sgB\,Y_{n-1}(\sfD)$ & $\sfQU\coloneqq$ & $\diamond\,\sLtinv\sfQ\,Y_{n-1}(\sgB)$ & Mixed terms$^{\ast}$\tabularnewline
\hline 
\multicolumn{1}{r}{} & \multicolumn{1}{l}{} & \multicolumn{1}{r}{} & \multicolumn{1}{l}{} & \multicolumn{1}{l}{}\tabularnewline[-4ex]
\multicolumn{4}{c}{$\hSector{\text{Geometric mean}}$} & \multicolumn{1}{l}{}\tabularnewline
\hline 
$\hSp=$ & $\gE^{\tr}\sEta\sLs\fE$ & $\hSp^{\tr}=$ & $\fE^{\tr}\sLs^{\tr}\sEta\gE$ & Spatial metric for $\hMet$\tabularnewline
\hline 
 & $\hSp=\gSp\sgB=\gSp\sgD^{-1}$ &  & $\hSp^{\tr}=\fSp\sfB=\fSp\sfD^{-1}$ & Note $\hSp=\hSp^{\tr}$\tabularnewline
\hline 
$\gSp^{-1}\fSp=$ & $\sgD^{-1}\sfD=\sgB\sfD$ & $\fSp^{-1}\gSp=$ & $\sfD^{-1}\sgD=\sfB\sgD$ & Follows from $\hSp=\hSp^{\tr}$\tabularnewline
\hline 
$\sLp=$ & $\sLs\sLv=\sLt\sLv$ & $\sLv=$ & $\gE\sgn=\fE\sfn$ & Boost parameter\tabularnewline
\hline 
$\sLt\coloneqq$ & $\big(1+\sLp^{{\scriptscriptstyle \tr}}\sEta\sLp\big)^{1/2}$ & $\sLt=$ & $\big(1-\sLv^{{\scriptscriptstyle \tr}}\sEta\sLv\big)^{-1/2}$ & Lorentz factor\tabularnewline
\hline 
$\sLs\coloneqq$ & $\big(\sI+\sLp\sLp^{\tr}\sEta\big)^{1/2}$ & $\sLs=$ & $\big(\sI-\sLv\sLv^{\tr}\sEta\big)^{-1/2}$ & Spatial part of LLT\tabularnewline
\hline 
\end{tabular}
\par\end{centering}
\egroup \vspace{1.2ex}
\centering{}%
\parbox[c][1\totalheight][t]{0.85\columnwidth}{%
$^{\ast}$ These variables are spanned by~ $\diamond\coloneqq\usignV\betaSum$.%
}
\end{table}

\begin{table}[h]
\vspace{5mm}\caption{\label{tab:dof}Number of degrees of freedom in GR and the HR bimetric
theory.}
\vspace{1ex}

\noindent \centering{}\bgroup\centering\renewcommand{\arraystretch}{1.4}%
\begin{tabular}{|c||c|c|c||c|c|c|}
\multicolumn{1}{c}{} & \multicolumn{3}{c}{GR} & \multicolumn{3}{c}{HR bimetric theory}\tabularnewline
\cline{2-7} 
\multicolumn{1}{c|}{} & gen.~case & $d=4$ & $d=3$ & gen.~case & $d=4$ & $d=3$\tabularnewline
\hline 
\hline 
Dynamical fields $(\gamma,K)$ & $d(d-1)$ & $12$ & $6$ & $2d(d-1)$ & $24$ & $12$\tabularnewline
\hline 
Constraint equations & $-d$ & $-4$ & $-3$ & $-(d+1)$ & $-5$ & $-4$\tabularnewline
\hline 
Bimetric conservation law & $0$ & $0$ & $0$ & $-1$ & $-1$ & $-1$\tabularnewline
\hline 
Gauge fixing & $-d$ & $-4$ & $-3$ & $-d$ & $-4$ & $-3$\tabularnewline
\hline 
\hline 
\# dof $=1/2$ of total & $d(d-3)/2$ & $2$ & $0$ & $d(d-2)-1$ & $7$ & $2$\tabularnewline
\hline 
\end{tabular}\egroup
\end{table}

\input{box-1.tex}

In conclusion to this section, we address one important condition
that the projected bimetric conservation law (\ref{eq:bim-prop-constr})
itself stays preserved under the dynamical evolution (i.e., that its
time derivative vanishes identically). Such a requirement imposes
a relation between the two lapses as shown in \cite{Hassan:2018mbl},
which gives the correct number of kinematical variables related to
the gauge freedom. To illustrate this, consider a $\beta_{1}$-model
where all the $\beta$-parameters vanish except $\beta_{1}$; then,
$\sgU=\sLtinv\sI$, $\sfU=\sfD$, and the preservation of (\ref{eq:bim-prop-constr})
requires,
\begin{equation}
\partial_{t}\big(\uisignK\sLtinv\gK+\sgn^{i}\partial_{i}\sLtinv\signK\tud{\sfD}ij\tud{\fK}ji-\tud{\sfD}ij\fD_{i}\sfn^{j}\big)=0.\label{eq:beta1-ex}
\end{equation}
The time derivative in (\ref{eq:beta1-ex}) can be resolved using
the evolution equations. For example, we can determine $\gLapse$
in terms of the other fields after setting $\hShift=0$ and $\fLapse=1$.
By a more general analysis, it can be shown that necessarily $\fLapse/\gLapse=W$,
where $W$ is a spatial scalar field determined from the other fields
\cite{Hassan:2018mbl}.\footnote{As a consequence, some combination of $(\gLapse,\fLapse)$ can be
taken as a primary kinematic variable; e.g., it can be the lapse of
$\hMet$ where $\hLapse^{2}=\gLapse\fLapse\sLtinv=\gLapse^{2}W\sLtinv$.
Choosing $\hLapse^{2}\sLt$ as an independent field (which can be
gauge fixed), one can define the lapses of $\gMet$ and $\fMet$ relative
to $\hMet$ as $\fLapse^{2}=\hLapse^{2}\sLt W$ and $\gLapse^{2}=\hLapse^{2}\sLt/W$.} The lapse ratio $\fLapse/\gLapse$ for the spherically symmetric
case is calculated in \cite{Kocic:2019zdy}. Note that the preservation
of the equation (\ref{eq:bim-prop-constr}) is sometimes referred
to as the stability condition for the secondary constraint in the
context of the Hamiltonian formalism \cite{Alexandrov:2012yv,Alexandrov:2013rxa,Hassan:2018mbl}.
However, the preservation and the stable propagation of the constraints
are usually considered as two distinct concepts \cite{Frittelli:1996nj},
where one assumes that a quantity propagates in a stable manner if
it depends continuously on its initial values \cite{Courant:1962vol2}.
Such a treatment is studied in \cite{Kocic:2018yvr}.

\makeatletter
\long\def\@makecaption#1#2{%
  \vskip\abovecaptionskip
  \sbox\@tempboxa{\small #1. #2}%
  \ifdim \wd\@tempboxa >\hsize
    \small #1. #2\par
  \else
    \global \@minipagefalse
    \hb@xt@\hsize{\hfil\box\@tempboxa\hfil}%
  \fi
  \vskip\belowcaptionskip}
\makeatother

%% file: box-1.tex
%%%%%%%%%%%%%%%%%%%%%%%%%%%%%%%%%%%%%%%%%%%%%%%%%%%%%%%%%%%%%%%%%%%%%%%%%%%%%

\input{macros.tex}
\newcommand{\eqMg}{%
\begin{align*}
\gMet_{\mu\nu} & =\left(\begin{array}{c|c}
1\, & \,0\\
\hline \gShift^{k} & \,\delta_{i}^{k}
\end{array}\right)^{\!\!\tr}\!\!\left(\begin{array}{c|c}
-\gLapse^{2}\, & \,0\\
\hline 0 & \,\gSp_{kl}
\end{array}\right)\!\!\left(\begin{array}{c|c}
1\, & \,0\\
\hline \gShift^{l} & \,\delta_{j}^{l}
\end{array}\right)\!.
\end{align*}
}

\newcommand{\eqMf}{%
\begin{align*}
\fMet & =\left(\begin{array}{c|c}
-\fLapse^{2}+\fShift^{k}\fSp_{kl}\fShift^{l}\, & \,\fShift^{k}\fSp_{kj}\\
\hline \fSp_{il}\fShift^{l}\, & \,\fSp_{ij}
\end{array}\right).
\end{align*}
}

\newcommand{\eqEg}{%
\begin{align*}
\partial_{t}\gSp_{ij} & =\Lie_{\hShift}\gSp_{ij}+\Lie_{\gLapse\sgn}\gSp_{ij}\signK2\gLapse\gK_{ij},\vphantom{\bigg|}\\
\partial_{t}\gK_{ij} & =\Lie_{\hShift}\gK_{ij}+\Lie_{\gLapse\sgn}\gK_{ij}\signK\gD_{i}\gD_{j}\gLapse\\
 & \isignK\gLapse\,\big[\gR_{ij}-2\gK_{ik}\gK^{k}{}_{j}+\gK\gK_{ij}\big]\\
 & \signKV\gLapse\gKappa\big[\gSp_{ik}\tud{\gJota}kj-\tfrac{1}{d-2}\gSp_{ij}(\gJota-\grho)\big].
\end{align*}
}

\newcommand{\eqEf}{%
\begin{align*}
\partial_{t}\fSp_{ij} & =\Lie_{\hShift}\fSp_{ij}-\Lie_{\fLapse\sfn}\fSp_{ij}\signK2\fLapse\fK_{ij},\vphantom{\bigg|}\\
\partial_{t}\fK_{ij} & =\Lie_{\hShift}\fK_{ij}-\Lie_{\fLapse\sfn}\fK_{ij}\signK\fD_{i}\fD_{j}\fLapse\\
 & \isignK\fLapse\,\big[\fR_{ij}-2\fK_{ik}\fK^{k}{}_{j}+\fK\fK_{ij}\big]\\
 & \signKV\fLapse\fKappa\big[\fSp_{ik}\tud{\fJota}ij-\tfrac{1}{d-2}\fSp_{ij}(\fJota-\frho)\big].
\end{align*}
}

\newcommand{\eqConstr}{%
\begin{alignat*}{2}
\text{Scalar constraints:} & \quad & 0 & =\gR+\gK^{2}-\gK_{ij}\gK^{ij}\isignV2\gKappa\,\grho,\\
 &  & 0 & =\fR+\fK^{2}-\fK_{ij}\fK^{ij}\isignV2\fKappa\,\frho,\\
\text{Vector constraints:} &  & 0 & =\gD_{k}\gK^{k}{}_{i}-\gD_{i}\gK\signKV\gKappa\,\gjota_{i},\vphantom{\stackrel{{\scriptscriptstyle (KV)}}{+}}\\
 &  & 0 & =\fD_{k}\fK^{k}{}_{i}-\fD_{i}\fK\signKV\fKappa\,\fjota_{i},\\
 &  & 0 & =\gKappainv\det\gE\,\big(\gD_{k}\gK^{k}{}_{i}-\gD_{i}\gK\big)+\fKappainv\det\fE\,\big(\fD_{k}\fK^{k}{}_{i}-\fD_{i}\fK\big),\vphantom{\stackrel{{\scriptscriptstyle (KV)}}{+}}\\
\text{Conservation law:} &  & 0 & =\tud{\sgU}ij\big(\gD_{i}\sgn^{j}\signK\tud{\gK}ij\big)+\tud{\sfU}ij\big(\fD_{i}\sfn^{j}\isignK\tud{\fK}ji\big)-\gD_{i}\big[\tud{\sgU}ij\sgn^{j}\big].
\end{alignat*}
}

\newcommand{\eqVgf}{%
\begin{alignat*}{2}
\grho & =-e_{n}(\sgB)=\sgn^{\tr}\gjota-\sgV, & \frho & =-\sLt e_{n-1}(\sfD)\det(\gE\fE^{-1}),\\
\gjota & =-\gSp\sgQ\sfU\sgn, & 0 & =\fjota+\gjota\det(\gE\fE^{-1}),\\
\gLapse\gJota & =\gLapse(\sgV-\sgQ\sfU)+\fLapse\sgU, & \fLapse\fJota & =[\fLapse(\sfV-\sfQ\sgU)+\gLapse\sfU]\det(\gE\fE^{-1}),\\
\gJota-\grho & =(d-n)V, & \qquad\fJota-\frho & =nV\det(\gE\fE^{-1}).
\end{alignat*}
}

\newcommand{\eqDict}{%
\begin{align*}
\gSp & \coloneqq\gE^{{\scriptscriptstyle \tr}}\sEta\gE, & \gSp^{-1}\fSp & =\sgD^{-1}\sfD=\sgB\sfD. & \fSp & \coloneqq\fE^{{\scriptscriptstyle \tr}}\sEta\fE, & \fSp^{-1}\gSp & =\sfD^{-1}\sgD=\sfB\sgD,\\
\gShiftVec & \coloneqq\hShift+\gLapse\gE^{-1}\sLv, & \sgD & \coloneqq\fE^{-1}\sLs^{-1}\gE, & \fShiftVec & \coloneqq\hShift-\fShift\fE^{-1}\sLv, & \sfD & \coloneqq\gE^{-1}\sLs^{-1}\fE,\\
\sgn & \coloneqq\gE^{-1}\sLv=\sLt\sfD\sfn, & \sgB & \coloneqq\sgD^{-1}=\sfD\sfQ, & \sfn & \coloneqq\fE^{-1}\sLv=\sLt\sgD\sgn, & \sfB & \coloneqq\sfD^{-1}=\sgD\sgQ,\\
\sgQ & \coloneqq\gE^{-1}\sLs^{2}\gE, & \sgV & \coloneqq e_{n}(\sfD), & \sfQ & \coloneqq\fE^{-1}\sLs^{2}\fE, & \sfV & \coloneqq\sLtinv e_{n-1}(\sgB),\\
\sgU & \coloneqq\sLtinv Y_{n-1}(\sgB), & \sfQU & =\sLtinv\sfQ\,Y_{n-1}(\sgB), & \sfU & \coloneqq\sfD\,Y_{n-1}(\sfD), & \sgQU & =\sgB\,Y_{n-1}(\sfD).
\end{align*}
\vspace{-6ex}
\begin{align*}
\text{Sign conventions:} & \quad\mathcal{L}_{\gMet,\fMet}^{\mathrm{int}}=\signV\betaSum e_{n}(S),\quad\gK_{ij}=\signK\tfrac{1}{2}\Lie_{\vec{n}_{\gMet}}\gSp_{ij}.\vphantom{\big|}\\
\text{Lorentz boost:} & \quad\sLs^{2}\coloneqq\sI+\sLp\sLp^{\tr}\sEta,\quad\sLt^{2}\coloneqq1+\sLp^{{\scriptscriptstyle \tr}}\sEta\sLp,\quad\sLp=\sLs\sLv=\sLt\sLv.\vphantom{\big|}\\
\text{Symmetry relations:} & \quad\gSp F(\sgB)=F(\sgB)^{\tr}\gSp,\quad\fSp F(\sfD)=F(\sfD)^{\tr}\fSp\quad(\text{for any }F).\vphantom{\big|}\\
Y_{n}(A)\coloneqq{\textstyle \sum_{k=0}^{n}}(-1)^{n+k} & e_{n}(A)\,A^{n-k}=\frac{\partial e_{n+1}(A)}{\partial A^{\tr}},\qquad Y_{n}(A)=e_{n}(A)\,I-A\,Y_{n-1}(A).\vphantom{\big|}
\end{align*}
}

\newcommand{\mytzTitle}[2]{%
  \begin{scope}
    \clip (#1.north west) rectangle ($(#1.north east)+(0,-7)$); 
    \draw [anchor=north west, very thin, rounded corners=5pt,
        minimum height=15, fill=black!80!blue!80] 
      (#1.north west) rectangle (#1.south east);
    \node [anchor=west,inner sep=5pt,text=white,
        font={\sf\fontsize{11pt}{0}\selectfont\bfseries}] 
      at ($(#1.north west)+(0,-3.5)$) 
      {#2};
  \end{scope}
}

\newcommand{\mytzEq}[3]{%
  \path let \p1 = ($(#1.east)-(#1.west)-(1,0)$) in
    node [anchor=center] 
      at ($(#1.center) + (0,#2)$) 
      {\begin{minipage}{\x1}#3\end{minipage}};
}

\begin{bbox}[H]\vspace{0ex}\hspace{-5mm}
\begin{tikzpicture}[x=1mm,y=1mm,node distance=1mm]

  \tikzset{R/.style={minimum height=#1mm}}
  \tikzset{C/.style={minimum width=#1mm, text width=#1mm-4mm}}
  \tikzset{boxB/.style={draw, anchor=north west, rectangle, 
      very thin,rounded corners=5pt, inner sep=5pt, C=78,
       color=gray!50}}

  \node [boxB,R=7,C=155,color=white,align=center] (TT) {\color{black}\textbf{Box 1.} \sf\bfseries The HR vacuum field equations in standard $\boldsymbol{N}$+1 form.\label{box-1}};

  %\node [below=of TT.south west,boxB,R=16] (Mg) {};
  %\node [right=of Mg.north east,boxB,R=16] (Mf) {};
  \node [below=of TT.south west,boxB,R=44,fill=gray!8] (Eg) {};
  \node [right=of Eg.north east,boxB,R=44,fill=gray!8] (Ef) {};
  \node [below=of Eg.south west,boxB,R=58,C=157,fill=gray!8] (Cs) {};
  \node [below=of Cs.south west,boxB,R=37,C=157] (Vgf) {};
  \node [below=of Vgf.south west,boxB,R=76,C=157] (Dict) {};

  \draw [anchor=north west, very thin, rounded corners=7pt] 
     ($(TT.north west)+(-1,1)$) rectangle ($(Dict.south east)+(1,-1)$);

  %\mytzEq{Mg}{2}{\eqMg}
  %\mytzEq{Mf}{2}{\eqMf}
  \mytzEq{Eg}{-1}{\eqEg}
  \mytzEq{Ef}{-1}{\eqEf}
  \mytzEq{Vgf}{-1}{\eqVgf}
  \mytzEq{Cs}{-1}{\eqConstr}
  \mytzEq{Dict}{-1}{\eqDict}

  \mytzTitle{Eg}{Evolution equations for $\boldsymbol{(\gamma,K)}$}
  \mytzTitle{Ef}{Evolution equations for $\boldsymbol{(\varphi,\tilde K)}$}
  \mytzTitle{Cs}{Constraint equations and the projection of the bimetric conservation law}
  \mytzTitle{Vgf}{Projections of $\boldsymbol{V_g}$ and $\boldsymbol{V_f}$ 
     for $\boldsymbol{V=e_n(S) = \mathcal{V} + M N^{-1} \widetilde{\mathcal{V}}}$}
  \mytzTitle{Dict}{Dictionary}

\end{tikzpicture}
\end{bbox}

%% file: sec-40.tex
\section{Spherically symmetric spacetimes}

\label{sec:ssym}

We now investigate the special case where the two metric sectors share
the same spherical symmetry. One reason is that the bimetric field
equations in spherical symmetry reduce to 1+1 dimensions and are simpler
to pose and solve. For instance, all the shift vectors become scalars
and we do not need to explicitly determine the spatial rotation $\sRs$
from (\ref{eq:sim-fE}). Also, the reduced 1+1 equations can serve
as a toy model for experimenting with different gauge choices and
slicings. As a physical motivation, many important aspects of gravitational
collapse can be studied in spherical symmetry, for example, black
hole or structure formation and growth. Finally, when treated numerically,
the spherical equations can be solved faster, and with higher accuracy
\cite{Baumgarte:2010numerical}. Hence, treating problems in spherical
symmetry seems as a good starting point for the bimetric numerical
relativity.

Let $d=4$. The general form of the metrics in spherical polar coordinates
reads \cite{Shibata:2015nr},\bSe\label{eq:ssym-gf}
\begin{align}
\gMet & =-\gAlpha^{2}\dd t^{2}+\gEA^{2}(\dd r+\gBeta\,\dd t)^{2}+\gEB^{2}(\dd\theta^{2}+\sin^{2}\theta\,\dd\phi^{2}),\\
\fMet & =-\fAlpha^{2}\dd t^{2}+\fEA^{2}(\dd r+\fBeta\,\dd t)^{2}+\fEB^{2}(\dd\theta^{2}+\sin^{2}\theta\,\dd\phi^{2}),
\end{align}
\eSe where $\gAlpha$ and $\fAlpha$ are the lapse functions, and
$(\gEA,\gEB,\fEA,\fEB)$ denote the nontrivial components of the spatial
vielbeins such that $\gSp_{rr}\coloneqq\gEA^{2}$, $\gSp_{\theta\theta}\coloneqq\gEB^{2}$,
$\fSp_{rr}\coloneqq\fEA^{2}$, and $\fSp_{\theta\theta}\coloneqq\fEB^{2}$.
The radial components of the shifts are parametrized by the mean shift
$\hShift$ and the radial separation $\sLp$,
\begin{equation}
\gBeta\coloneqq\hShift+\gAlpha\gEA^{-1}\sLv,\quad\fBeta\coloneqq\hShift-\fAlpha\fEA^{-1}\sLv,\quad\sLv\coloneqq\sLp\sLtinv,\quad\sLt\coloneqq(1+\sLp^{2})^{1/2}.
\end{equation}
In addition, we have the components of the extrinsic curvature,
\begin{equation}
\gK_{1}\eqqcolon\tud{\gK}rr,\quad\gK_{2}\eqqcolon\tud{\gK}{\theta}{\theta}=\tud{\gK}{\phi}{\phi},\quad\fK_{1}\eqqcolon\tud{\fK}rr,\quad\fK_{2}\eqqcolon\tud{\fK}{\theta}{\theta}=\tud{\fK}{\phi}{\phi}.
\end{equation}
All these variables are functions of $(t,r)$ to be solved for in
general.

For the following, we introduce $\sER\coloneqq\fEB/\gEB$ and define
a partial span by $\{\beta_{n}\}$,
\begin{equation}
\left\langle \sER\right\rangle _{k}^{n}\coloneqq\usignV\betaScale\sum_{i=0}^{n}\mbinom ni\beta_{i+k}\sER^{i},\quad\ \left\langle \sER\right\rangle _{k}^{n}=\left\langle \sER\right\rangle _{k}^{n-1}+\sER\left\langle \sER\right\rangle _{k+1}^{n-1},\ \left\langle \sER\right\rangle _{k}^{0}=\usignV\betaScale\beta_{k}.
\end{equation}
The nonzero components of the projections of the bimetric stress-energy
tensor $V_{\gMet}$ are,\bSe\label{eq:ssym-g-se}\vspace{-1ex}
\begin{align}
\grho & =-\Bigg[\left\langle \sER\right\rangle _{0}^{2}+\sLt\frac{\fEA}{\gEA}\left\langle \sER\right\rangle _{1}^{2}\Bigg],\qquad\gjota_{r}=-\sLp\fEA\left\langle \sER\right\rangle _{1}^{2},\\
\gJota_{1} & =\left\langle \sER\right\rangle _{0}^{2}+\Bigg[\frac{1}{\sLt}\bigg(\frac{\fAlpha}{\gAlpha}+\frac{\fEA}{\gEA}\bigg)-\sLt\frac{\fEA}{\gEA}\Bigg]\left\langle \sER\right\rangle _{1}^{2},\\
\gJota_{2} & =\left\langle \sER\right\rangle _{0}^{1}+\frac{\fAlpha\fEA}{\gAlpha\gEA}\left\langle \sER\right\rangle _{1}^{2}+\frac{1}{\sLt}\bigg(\frac{\fAlpha}{\gAlpha}+\frac{\fEA}{\gEA}\bigg)\left\langle \sER\right\rangle _{1}^{1},
\end{align}
\eSe where $\gJota_{1}\coloneqq\tud{\gJota}rr$, $\gJota_{2}\coloneqq\tud{\gJota}{\theta}{\theta}=\tud{\gJota}{\phi}{\phi}$,
$\gJota=\gJota_{1}+2\gJota_{2}$. Similarly for $V_{\fMet}$ we have,\bSe
\label{eq:ssym-f-se}\vspace{-1ex}
\begin{align}
\frho & =-\Bigg[\left\langle \sER\right\rangle _{2}^{2}+\sLt\frac{\gEA}{\fEA}\left\langle \sER\right\rangle _{1}^{2}\Bigg]\frac{1}{\sER^{2}},\qquad\fjota_{r}=\sLp\gEA\left\langle \sER\right\rangle _{1}^{2}\frac{1}{\sER^{2}},\\
\fJota_{1} & =\Bigg\{\left\langle \sER\right\rangle _{2}^{2}+\Bigg[\frac{1}{\sLt}\bigg(\frac{\gAlpha}{\fAlpha}+\frac{\gEA}{\fEA}\bigg)-\sLt\frac{\gEA}{\fEA}\Bigg]\left\langle \sER\right\rangle _{1}^{2}\Bigg\}\frac{1}{\sER^{2}},\\
\fJota_{2} & =\Bigg\{\left\langle \sER\right\rangle _{3}^{1}+\frac{\gAlpha\gEA}{\fAlpha\fEA}\left\langle \sER\right\rangle _{1}^{1}+\frac{1}{\sLt}\bigg(\frac{\gAlpha}{\fAlpha}+\frac{\gEA}{\fEA}\bigg)\left\langle \sER\right\rangle _{2}^{1}\Bigg\}\frac{1}{\sER},
\end{align}
\eSe where $\fJota_{1}\coloneqq\tud{\fJota}rr$, $\fJota_{2}\coloneqq\tud{\fJota}{\theta}{\theta}=\tud{\fJota}{\phi}{\phi}$,
and $\fJota=\fJota_{1}+2\fJota_{2}$. The \nPlusOne{} variables from
Table \ref{tab:vars} are given for the case of spherical symmetry
in appendix \ref{app:ssym-vars}.

The scalar constraints (\ref{eq:g-cc-scalar})\textendash (\ref{eq:f-cc-scalar})
are,\bSe\label{eq:ssym-cc}\vspace{-0.5ex}
\begin{align}
(2\gK_{1}+\gK_{2})\gK_{2}+\frac{1}{\gEA^{2}}\bigg(\frac{\gEA^{2}}{\gEB^{2}}+2\frac{\partial_{r}\gEA}{\gEA}\frac{\partial_{r}\gEB}{\gEB}-\frac{(\partial_{r}\gEB)^{2}}{\gEB^{2}}-2\frac{\partial_{r}^{2}\gEB}{\gEB}\bigg) & =\gKappa\grho,\\
(2\fK_{1}+\fK_{2})\fK_{2}+\frac{1}{\fEA^{2}}\bigg(\frac{\fEA^{2}}{\fEB^{2}}+2\frac{\partial_{r}\fEA}{\fEA}\frac{\partial_{r}\fEB}{\fEB}-\frac{(\partial_{r}\fEB)^{2}}{\fEB^{2}}-2\frac{\partial_{r}^{2}\fEB}{\fEB}\bigg) & =\fKappa\frho.
\end{align}
The vector constraints (\ref{eq:g-cc-vector})\textendash (\ref{eq:f-cc-vector})
are, 
\begin{align}
(\gK_{1}-\gK_{2})\frac{\partial_{r}\gEB}{\gEB}-\partial_{r}\gK_{2} & =\uisignK\frac{1}{2}\,\gKappa\gjota_{r},\\
(\fK_{1}-\fK_{2})\frac{\partial_{r}\fEB}{\fEB}-\partial_{r}\fK_{2} & =\uisignK\frac{1}{2}\,\fKappa\fjota_{r}.\label{eq:ssym-eq-j}
\end{align}
The last two equations can be recombined using (\ref{eq:f-cc-j-def}),
\begin{equation}
\fKappa\fEA\gEB\big(\gK_{1}\partial_{r}\gEB-\gK_{2}\partial_{r}\gEB-\gEB\partial_{r}\gK_{2}\big)+\gKappa\gEA\fEB\big(\fK_{1}\partial_{r}\fEB-\fK_{2}\partial_{r}\fEB-\fEB\partial_{r}\fK_{2}\big)=0.
\end{equation}
\eSe The radial shift separation can be determined from (\ref{eq:ssym-eq-j})
as,
\begin{equation}
\sLp=\usignK\frac{2}{\gKappa\fEA\gEB\left\langle \sER\right\rangle _{1}^{2}}\big(\gK_{1}\partial_{r}\gEB-\gK_{2}\partial_{r}\gEB-\gEB\partial_{r}\gK_{2}\big).\label{eq:ssym-eq-p}
\end{equation}
The projection (\ref{eq:bim-prop-constr}) of the bimetric conservation
law reads,
\begin{gather}
\fEA\Big(\fK_{1}\left\langle \sER\right\rangle _{1}^{2}+2\fK_{2}\sER\left\langle \sER\right\rangle _{2}^{1}\Big)-\gEA\Big(\gK_{1}\left\langle \sER\right\rangle _{1}^{2}+2\gK_{2}\left\langle \sER\right\rangle _{1}^{1}\Big)+2\gEA\fK_{2}\sLt\sER\left\langle \sER\right\rangle _{1}^{1}-2\fEA\gK_{2}\sLt\left\langle \sER\right\rangle _{2}^{1}\nonumber \\
\isignK2\sLp\bigg(\left\langle \sER\right\rangle _{1}^{1}\frac{\gEA}{\fEA}\frac{\partial_{r}\fEB}{\gEB}+\left\langle \sER\right\rangle _{2}^{1}\frac{\fEA}{\gEA}\frac{\partial_{r}\gEB}{\gEB}\bigg)\isignK\sLtinv\left\langle \sER\right\rangle _{1}^{2}\partial_{r}\sLp=0,\label{eq:ssym-cc2}
\end{gather}
where $\sLp$ can be eliminated using (\ref{eq:ssym-eq-p}). The evolution
equations for the spatial metrics are,\bSe\label{eq:ssym-evol-1}
\begin{alignat}{2}
\partial_{t}\gEA & =\usignK\gAlpha\gEA\gK_{1}+\partial_{r}(\hShift\gEA+\gAlpha\sLv), & \qquad\partial_{t}\gEB & =\usignK\gAlpha\gEB\gK_{2}+\big(\hShift+\gAlpha\gEA^{-1}\sLv\big)\partial_{r}\gEB,\\
\partial_{t}\fEA & =\usignK\fAlpha\fEA\fK_{1}+\partial_{r}(\hShift\fEA-\fAlpha\sLv), & \partial_{t}\fEB & =\usignK\fAlpha\fEB\fK_{2}+\big(\hShift-\fAlpha\fEA^{-1}\sLv\big)\partial_{r}\fEB.
\end{alignat}
\eSe The evolution equations for the extrinsic curvatures are,\bSe\label{eq:ssym-evol-2}
\begin{align}
\partial_{t}\gK_{1} & =\big(\hShift+\gAlpha\gEA^{-1}\sLv\big)\partial_{r}\gK_{1}\isignK\gAlpha\gK_{1}\big(\gK_{1}+2\gK_{2}\big)\signK\gAlpha\gKappa\Big\{\,\gJota_{1}-\frac{1}{2}(\gJota-\grho)\,\Big\}\nonumber \\
 & \qquad\isignK\bigg(\frac{\partial_{r}\gAlpha}{\gEA^{2}}\frac{\partial_{r}\gEA}{\gEA}-\frac{\partial_{r}^{2}\gAlpha}{\gEA^{2}}+2\frac{\gAlpha}{\gEA^{2}}\frac{\partial_{r}\gEA}{\gEA}\frac{\partial_{r}\gEB}{\gEB}-2\frac{\gAlpha}{\gEA^{2}}\frac{\partial_{r}^{2}\gEB}{\gEB}\bigg),\\
\partial_{t}\fK_{1} & =\big(\hShift-\fAlpha\fEA^{-1}\sLv\big)\partial_{r}\fK_{1}\isignK\fAlpha\fK_{1}\big(\fK_{1}+2\fK_{2}\big)\signK\fAlpha\fKappa\Big\{\,\fJota_{1}-\frac{1}{2}(\fJota-\frho)\,\Big\}\nonumber \\
 & \qquad\isignK\bigg(\frac{\partial_{r}\fAlpha}{\fEA^{2}}\frac{\partial_{r}\fEA}{\fEA}-\frac{\partial_{r}^{2}\fAlpha}{\fEA^{2}}+2\frac{\fAlpha}{\fEA^{2}}\frac{\partial_{r}\fEA}{\fEA}\frac{\partial_{r}\fEB}{\fEB}-2\frac{\fAlpha}{\fEA^{2}}\frac{\partial_{r}^{2}\fEB}{\fEB}\bigg),\\
\partial_{t}\gK_{2} & =\big(\hShift+\gAlpha\gEA^{-1}\sLv\big)\partial_{r}\gK_{2}\isignK\gAlpha\gK_{2}\big(\gK_{1}+2\gK_{2}\big)\signK\gAlpha\gKappa\Big\{\,\gJota_{2}-\frac{1}{2}(\gJota-\grho)\,\Big\}\nonumber \\
 & \qquad\isignK\bigg(\frac{\gAlpha}{\gEB^{2}}-\frac{\partial_{r}\gAlpha}{\gEA^{2}}\frac{\partial_{r}\gEB}{\gEB}+\frac{\gAlpha}{\gEA^{2}}\frac{\partial_{r}\gEA}{\gEA}\frac{\partial_{r}\gEB}{\gEB}-\frac{\gAlpha}{\gEA^{2}}\frac{(\partial_{r}\gEB)^{2}}{\gEB^{2}}-\frac{\gAlpha}{\gEA^{2}}\frac{\partial_{r}^{2}\gEB}{\gEB}\bigg),\\
\partial_{t}\fK_{2} & =\big(\hShift-\fAlpha\fEA^{-1}\sLv\big)\partial_{r}\fK_{2}\isignK\fAlpha\fK_{2}\big(\fK_{1}+2\fK_{2}\big)\signK\fAlpha\fKappa\Big\{\,\fJota_{2}-\frac{1}{2}(\fJota-\frho)\,\Big\}\nonumber \\
 & \qquad\isignK\bigg(\frac{\fAlpha}{\fEB^{2}}-\frac{\partial_{r}\fAlpha}{\fEA^{2}}\frac{\partial_{r}\fEB}{\fEB}+\frac{\fAlpha}{\fEA^{2}}\frac{\partial_{r}\fEA}{\fEA}\frac{\partial_{r}\fEB}{\fEB}-\frac{\fAlpha}{\fEA^{2}}\frac{(\partial_{r}\fEB)^{2}}{\fEB^{2}}-\frac{\fAlpha}{\fEA^{2}}\frac{\partial_{r}^{2}\fEB}{\fEB}\bigg).
\end{align}
\eSe The algebraic symmetry between the two sectors is obvious. The
set of 3+1 equations in each bimetric sector is directly comparable
to GR \cite{Baumgarte:2010numerical,Shibata:2015nr}.

As a sanity check, let us decouple $\gMet$ and $\fMet$ by setting
$\betaScale\beta_{0}=\Lambda_{\gMet}$, $\betaScale\beta_{4}=\Lambda_{\fMet}$,
and $\beta_{1}=\beta_{2}=\beta_{3}=0$. We get the stress-energy components
for the cosmological constants,
\begin{equation}
\grho=\uisignV\Lambda_{\gMet},\ \gjota_{r}=0,\ \gJota_{1}=\gJota_{2}=\usignV\Lambda_{\gMet},\quad\frho=\uisignV\Lambda_{\fMet},\ \fjota_{r}=0,\ \fJota_{1}=\fJota_{2}=\usignV\Lambda_{\fMet}.\label{eq:ssym-cconst}
\end{equation}
Here we also have $\sgU=0$ and $\sfU=0$, so the bimetric conservation
law becomes automatically satisfied. Notice the implications of choosing
the $V$-sign convention (inside $\left\langle \sER\right\rangle _{k}^{n}$)
on the bimetric stress-energy projections (\ref{eq:ssym-g-se}), (\ref{eq:ssym-f-se}),
and (\ref{eq:ssym-cconst}).\newpage

As another sanity check, we can compare the 3+1 split to the existing
cosmological and black hole solutions exhibiting spherical symmetry
in the HR bimetric theory. We start with the derivation of cosmological
solutions \cite{vonStrauss:2011mq} based on the Fried\-mann\textendash Le\-ma\^i\-tre\textendash Rob\-ert\-son\textendash Walk\-er
ansatz. Therein, the lapse of $\fMet$ is denoted as $X$, while the
lapse of $\gMet$ is fixed to one, so the ratio of the two lapses
reads $W=\fAlpha/\gAlpha=X$. The constraint equations reduce to the
equation (2.12) in \cite{vonStrauss:2011mq}. Moreover, the equation
(2.13) fixes the ratio between the lapses and corresponds to the preservation
of the bimetric conservation law condition. On the other hand, for
a comparison with the black hole solutions, let us consider a choice
of variables in the ``radial gauge'' $\gEB=r$ where we take $\sER=\fEB/\gEB$
as a primary field. Furthermore, we express the equations in terms
of two additional scalar functions $\sEtau=\fAlpha/\gAlpha$ and $\sESigma=\fEA/\gEA$
(which appear in the eigenvalues of the square root). These names
are complaint with the spherically symmetric ansatz from \cite{Torsello:2017cmz}.
 The constraints reduce to the fifth equation (23e) in \cite{Torsello:2017cmz}.
The algebraic condition for $\sEtau$ (25a) in \cite{Torsello:2017cmz}
fixes $\sEtau$ relative to the other fields, and roughly corresponds
to the preservation of the bimetric conservation law condition imposing
the ratio between the two lapses $\sEtau=\fAlpha/\gAlpha$. Note that
the most general form of the lapse ratio $\fAlpha/\gAlpha$ for the
spherically symmetric case is calculated in \cite{Kocic:2019zdy}.

The variables can be differently reparametrized to simplify the equations.
For example, one can use the exponential functions instead of $\gEA,\gEB,\fEA,\fEB$,
or introduce the densitized lapses $\gSector{\underbar{\ensuremath{\gAlpha}}}=\gAlpha\gEA^{-1}$
and $\fSector{\underbar{\ensuremath{\fAlpha}}}=\fAlpha\fEA^{-1}$
which nicely appear as the factors in the shift separations. Also,
one option is to do a spatial gauge fixing and work in a zero shift
$\hShift=0$.

The next step would be to study the required conditions for the gauge
choice in general. For example, the lapse function can be determined
from the maximal slicing which is done with respect to the geometric
mean metric, functioning as a common ``singularity avoiding'' slicing
condition for both sectors. However, there are many other choices
for gauge fixing that may give the condition for the lapse function
(the time slicing condition), and the condition for the shift vector
(the spatial gauge condition). Posing such conditions and then solving
the spherical equations will be treated elsewhere.

\paragraph*{Degrees of freedom in spherical symmetry.}

Here we determine the number of truly dynamical variables in the case
of spherical symmetry. In GR, the spherical reduction of the number
of degrees of freedom is reflected in the fact that we can obtain
a fully constrained system with no evolution equations, which is compliant
with Birkhoff's theorem. On the other hand, the constraint equations
in the HR theory cannot remove all of the dynamical degrees of freedom
after imposing the spherical symmetry. In particular, the spherical
symmetry reduction  does not suppress the propagating degrees of
freedom associated with the monopole radiation (contrary to Birkhoff's
theorem). To see this, let us start with the field inventory $(\sLp,\hShift,\gAlpha,\gEA,\gEB,\gK_{1},\gK_{2},\fAlpha,\fEA,\fEB,\fK_{1},\fK_{2})$.
We also have eight evolution equations in (\ref{eq:ssym-evol-1})
and (\ref{eq:ssym-evol-2}), four constraint equations in (\ref{eq:ssym-cc}),
one conservation of the bimetric potential equation (\ref{eq:ssym-cc2}),
and one equation relating the two lapses. The gauge freedom can be
used to fix $\gAlpha$ and $\hShift$ eliminating one conjugate pair.
One of the constraint equations can be used to determine $\sLp$ (\ref{eq:ssym-eq-p}).
At the end, we are left with one dynamical conjugate pair governing
radial fluctuations in time.

%% file: sec-60.tex
\section{Summary and outlook}

\label{sec:discussion}

Various averaging operations on matrices have been of interest to
operator theorists, physicists, and statisticians for a long time
\cite{Bhatia2009:posdef}. Particularly intriguing has been the notion
of the geometric mean of positive definite matrices \cite{Ando:2004aa,Bhatia:2006aa,Bhatia2009:posdef}.

\begin{figure}
\noindent \centering{}\hspace{2mm}\includegraphics[scale=0.92]{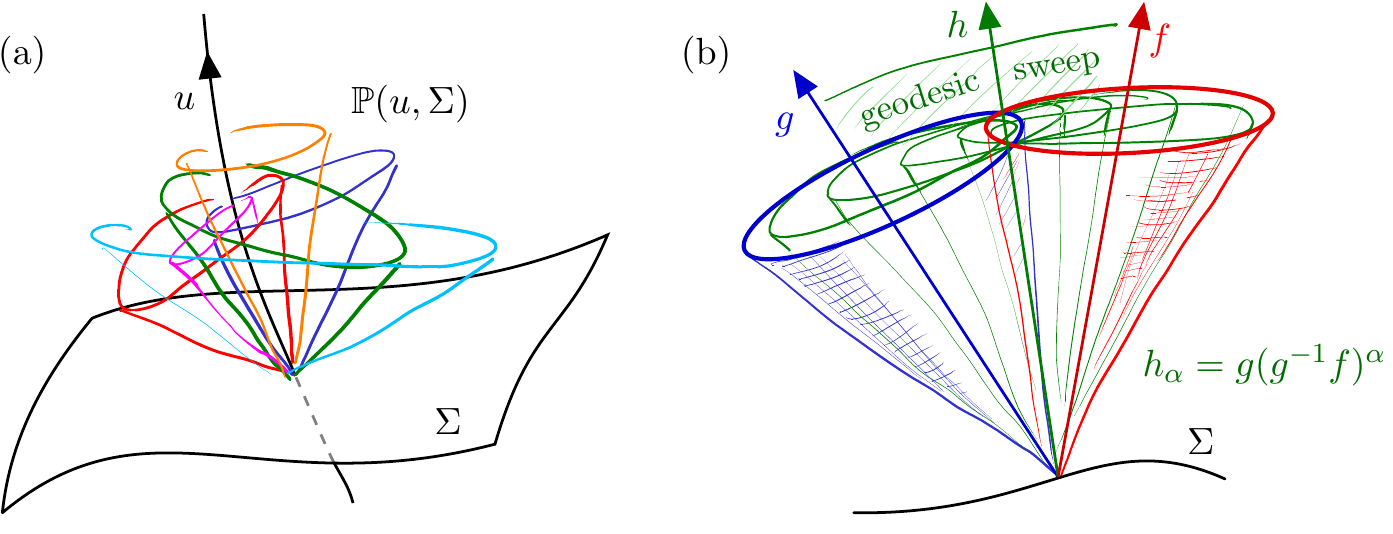}\vspace{-2ex}\caption{\label{fig:met-space}The metric space of metrics $\mathbb{P}(u,\Sigma)$,
and the geodesic $h_{\alpha}$ connecting $g$ and $f$.}
\end{figure}

Now, let us consider a space $\mathbb{P}(u,\Sigma)$ of all Lorentzian
metrics for which a vector field $u$ is future-pointing timelike
and a family of nonintersecting hypersurfaces $\{\Sigma\}$ is spacelike,
as illustrated in Figure \ref{fig:met-space}a. We assume that the
spacelike slices $\{\Sigma\}$ locally arise as the level surfaces
of a smooth scalar field $\tau$ called the time function. In particular,
the foliation is set out by a closed one-form $\Omega$, and since
$\Omega$ is closed, there locally exists a scalar field $\tau$ such
that $\Omega=\dd\tau$, where we also assume $u^{\mu}\Omega_{\mu}\ne0$
so that $u$ is never tangent to $\{\Sigma\}$.

The space $\mathbb{P}(u,\Sigma)$ can be endowed with a Riemannian
metric $\Delta$ measuring the distance between two metrics $g_{1}$
and $g_{2}$ (which are points in $\mathbb{P}$) defined as,
\begin{equation}
\Delta(g_{1},g_{2})\coloneqq\left\Vert \,\log(g_{1}^{-1}g_{2})\,\right\Vert ,\qquad\left\Vert X\right\Vert \coloneqq\Big(\,{\textstyle \sum_{\lambda\in\sigma[X]}}\lambda^{2}\,\Big){}^{1/2},\label{eq:gm-metric}
\end{equation}
where $\sigma[X]$ denotes the set of the eigenvalues (spectra) of
$X$. Strictly, the metric $\Delta$ can be defined only on a space
of positive definite symmetric matrices \cite{Lawson:2001aa}.\footnote{The Riemannian distance $\Delta$ arises as the metric associated
with arc length for the trace metric: $\dd s^{2}=\Tr(g^{-1}\dd g)^{2}$.
For a curve $t\mapsto g(t)$, the differential for computing length
is given by $(\dd s/\dd t)^{2}=$ $\Tr[g(t)^{-1}g^{\prime}(t)]^{2}$.
In dimension one, $\Delta(x,y)=\left|\log(y/x)\right|$ coincides
with the distance between $\log x$ and $\log y$, which arises as
the minimal arc length distance for the Riemannian metric $\dd s^{2}=t^{-2}\dd t^{2}$.} Here we extend such a definition to $\mathbb{P}(u,\Sigma)$ since
the conditions for the existence of the principal logarithm are the
same as for the existence of the principal square root \cite{Higham:2008}.
Consequently, the validity of the definition (\ref{eq:gm-metric})
comes from the theorem from \cite{Hassan:2017ugh}. This makes applicable
the considerations from \cite{Lawson:2001aa,Ando:2004aa,Bhatia:2006aa,Bhatia:2011aa}.
The set of spaces $\mathbb{P}(u,\Sigma)$ can be identified as a natural
habitat of the solutions in the HR bimetric theory. An open problem
is still a definition of the geometric mean of several Lorentzian
metrics.

An important observation is that any two points $g,f$ in $\mathbb{P}(u,\Sigma)$
can be joined by a unique geodesic for which a natural parametrization
is given by,
\begin{equation}
h_{\alpha}=g\op{\#}_{\alpha}f=g(g^{-1}f)^{\alpha},\qquad0\le\alpha\le1.\label{eq:h-geodesic}
\end{equation}
The geodesic can be illustrated by a null cone sweep, as shown in
Figure \ref{fig:met-space}b. In \cite{Torsello:2017ouh} it was shown
that, if any two metrics on a geodesic (\ref{eq:h-geodesic}) share
the same Killing vector field, then all the metrics on the geodesic
share the same isometry. The geometric mean (\ref{eq:gm-1}) is obviously
the midpoint of the geodesic (\ref{eq:h-geodesic}), which justifies
the expression that the null cone of $h$ is `in the middle' of the
null cones of $g$ and $f$. The midpoint is unique since $\Delta(g,f)=2\Delta(g,g\op{\#}f)$
and $\Delta(g,f)=2\Delta(f,g\op{\#}f)$. 

The geometric mean metric can be used, in principle, to measure a
mean curvature when adapting a singularity avoiding slicing. Namely,
having GR as the guideline, a simple choice of setting the time and
the spatial slicing conditions will not work. The selection of the
gauge conditions needs to be physically and geometrically motivated
so that the lapse and shift are adapted to be suitable for the corresponding
spacetime. In this sense, the parametrization based on the geometric
mean may be usable in gauge fixing. 

Note that any dynamical choice of the lapse and shift as functions
of the metric or first-order derivatives of the metric is allowed,
provided that the evolution of the bimetric constraints is well-posed.
Ensuring the stable propagation of the constraints is important since,
if the constraint evolution equations are not well-posed, the unphysical
modes (which are normally suppressed by the constraint equations)
will not be bounded but propagated as amplified by the free evolution.
The causal propagation of the bimetric constraints is studied in \cite{Kocic:2018yvr}.

The \nPlusOne{} form of bimetric equations derived here is compatible
with York's standard version that is used in GR. The applied procedure
is complementary to the one using the Hamiltonian formalism \cite{Hassan:2018mbl}.
The variables in the Hamiltonian formalism are evaluated by varying
the \nPlusOne{} form of the HR action, while the variables in this
work are derived by \nPlusOne{} projecting of the bimetric field
equations obtained from the HR action varied in a general form. As
shown, the \nPlusOne{} projection was straightforward because of
the duality between the two metrics in the HR theory. Both procedures
yield the same form of the secondary constraint (for a detailed comparison
of the variables see appendix \ref{app:rosetta}).

The 3+1 form of the equations for the spherically symmetric case was
given in section\,\ref{sec:ssym}. One of the applications is in
the study of gravitational collapse in spherical symmetry~\cite{Kocic:poly}.
Finally, further modification of the evolution equations is necessary
for achieving numerical stability (for instance, employing the BSSN
formalism \cite{Shibata:1995we,Baumgarte:1998te,Torsello:2019aaa}).

\acknowledgments

I am grateful to Anders Lundkvist for plenty of discussion on the
Hamiltonian formalism in bimetric relativity. It is also my pleasure
to thank Fawad Hassan, Francesco Torsello, Edvard M\"{o}rtsell, and
Marcus H\"{o}g\r{a}s for valuable discussions and reading the manuscript.
Special thanks are due to Ingemar Bengtsson for reviving my interest
in the geometric means of matrices.

%% file: sec-90.tex
\section{Projections of $\nabla_{\mu}\protect\tud X{\mu}{\nu}=0$}

\label{app:proj-nabla-T}

Let $\rho\coloneqq\prho[X]$, $j\coloneqq\pjota[X]$, and $J\coloneqq\pJota[X]$.
The operator $X$ can be, for instance, $\gTse$, $\gEinst-\gKappa\gTse$,
or $\gVse$. We summarize the derivations \cite{Gourgoulhon:2012trip}
inspired by the paper of Frittelli \cite{Frittelli:1996nj} (similar
expressions can also be found in \cite{Shibata:2015nr}). A direct
expansion yields,
\begin{align}
\gCD_{\mu}\tud X{\mu}{\alpha} & =\gCD_{\mu}\big(\tud J{\mu}{\alpha}+n^{\mu}j_{\alpha}+j^{\mu}n_{\alpha}+\rho n^{\mu}n_{\alpha}\big)\\
 & =\gCD_{\mu}\tud J{\mu}{\alpha}-\gK j_{\alpha}+n^{\mu}\gCD_{\mu}j_{\alpha}+\gCD_{\mu}j^{\mu}\,n_{\alpha}-j^{\mu}\gK_{\mu\alpha}-\gK\rho n_{\alpha}\nonumber \\
 & \qquad\qquad+\,\rho\gD_{\alpha}\log\gLapse+n^{\mu}\gCD_{\mu}\rho\,n_{\alpha}.
\end{align}
The projection of $\gCD_{\mu}\tud X{\mu}{\nu}=0$ along $n^{\nu}$
reads,
\begin{equation}
n^{\nu}\gCD_{\mu}\tud J{\mu}{\nu}+n^{\mu}n^{\nu}\gCD_{\mu}j_{\nu}-\gCD_{\mu}j^{\mu}\isignK\gK\rho-n^{\mu}\gCD_{\mu}\rho=0,
\end{equation}
which can be simplified to,
\begin{equation}
\frac{1}{\gLapse}\big(\partial_{t}-\Lie_{\gShiftVec}\rho\big)+\gD_{i}j^{i}\signK\gK\rho\signK\gK_{ij}J^{ij}+2j^{i}\gD_{i}\log\gLapse=0.
\end{equation}
Here we used $\gCD_{\mu}n_{\nu}=\isignK\gK_{\mu\nu}+n_{\mu}\gD_{\nu}\log\gLapse$
and,
\begin{gather}
n^{\nu}\gCD_{\mu}\tud J{\mu}{\nu}=-\tud J{\mu}{\nu}\gCD_{\mu}n^{\nu}=\tud J{\mu}{\nu}(\isignK\tud{\gK}{\nu}{\mu}+n_{\mu}\gD^{\nu}\log\gLapse)=\isignK\gK_{\mu\nu}J^{\mu\nu},\\
n^{\mu}n^{\nu}\gCD_{\mu}j_{\nu}=-j_{\nu}n^{\mu}\gCD_{\mu}n^{\nu}=-j_{\nu}\gD^{\nu}\log\gLapse.
\end{gather}
Also, note that for an arbitrary tangential vector $\qvf$ holds $\gCD_{\mu}\qvf^{\mu}=\gD_{i}\qvf^{i}+\qvf^{i}\gD_{i}\log\gLapse.$

Similarly, the projection $\gCD_{\mu}\tud X{\mu}{\nu}=0$ using $\tud{\Proj}{\nu}{\alpha}$
reads,
\begin{equation}
\tud{\Proj}{\nu}{\alpha}\gCD_{\mu}\tud J{\mu}{\nu}-\gK j_{\alpha}+\tud{\Proj}{\nu}{\alpha}n^{\mu}\gCD_{\mu}j_{\nu}-\gK_{\alpha\mu}j^{\mu}+\rho\gD_{\alpha}\log\gLapse=0,
\end{equation}
which can be simplified to,
\begin{equation}
\frac{1}{\gLapse}\big(\partial_{t}-\Lie_{\gShiftVec}j_{i}\big)+\gD_{j}\tud Jji+J_{ij}\gD^{j}\log\gLapse-\gK j_{i}+\rho\gD_{i}\log\gLapse=0.
\end{equation}

\section{Proof of Proposition~\ref{prop:param}}

\label{app:proof-prop-1}

Here we combine the results from the sections 3 and 5 of \cite{Hassan:2014gta},
the theorem from \cite{Hassan:2017ugh}, and one additional lemma
(given below). If $S$ is an arbitrary congruence, $\fMet=S^{\tr}\gMet S$,
then the symmetrization condition $\hMet=\gMet S=\hMet^{\tr}$ is
equivalent to the requirement that the congruence $S$ is symmetric
(self-adjoint), $S=S^{\prime}$ where $S^{\prime}\coloneqq\gMet^{-1}S^{\tr}\gMet$,
which is further equivalent to the fact that $S$ is a square root,
$\fMet=\gMet S^{2}$. In terms of vielbeins, where $\gMet=E(\gMet)^{\tr}\eta E(\gMet)$
and $\fMet=E(\fMet)^{\tr}\Lambda^{\tr}\eta\Lambda E(\fMet)$, the
symmetrization condition reads \cite{Deffayet:2012zc}, 
\begin{equation}
\hMet=\gMet S=E(\gMet)^{\tr}\eta\Lambda E(\fMet)=\hMet^{\tr},\label{eq:sym-cond}
\end{equation}
where $\Lambda$ is a residual overall local Lorentz transformation
(LLT). 

The relation between the square root $S$ and the Lorentz transformation
$\Lambda$ is just the change of basis $S=E(\gMet)^{-1}\eta\Lambda E(\fMet)$.
Before proceeding, we address one important issue.

\begin{lemma}\label{lemma:vielb}An arbitrary vielbein can be triangularized
by a local Lorentz transformation if and only if the apparent lapse
of the associated metric is real in a given coordinate chart.\end{lemma}

\noindent \emph{Proof}. Consider an arbitrary vielbein $E$ that
we want to triangularize by a LLT $\Lambda$,
\begin{equation}
E=\begin{pmatrix}e_{0} & e_{1}^{\tr}\\
e_{2} & e_{3}
\end{pmatrix},\qquad\Lambda E=\begin{pmatrix}\gLapse & 0\\
\gE\gShiftVec & \gE
\end{pmatrix},\qquad\Lambda=\begin{pmatrix}\sLt\, & \sLp^{\prime}\\
\sLp & \sLs
\end{pmatrix}\!\!\begin{pmatrix}1 & 0\\
0 & \sRs
\end{pmatrix}\!.
\end{equation}
The equation for the upper right component reads $e_{1}^{\tr}+\sLp^{\prime}\sRs e_{3}=0$,
which becomes $\sLv^{\prime}\sRs e_{3}=-e_{1}^{\tr}$ for $\sLv=\sLp\sLtinv$.
The vector $\sLv$ can be determined,
\begin{equation}
\sLv^{\prime}=\sLv^{\tr}\sEta=-e_{1}^{\tr}e_{3}^{-1}\sEta^{-1}\sRs^{\tr}\sEta,\qquad\sLv=-(e_{1}^{\tr}e_{3}^{-1}\sEta^{-1}\sRs^{\tr})^{\tr}=-\sRs\sEta^{-1}e_{3}^{-1,\tr}e_{1}.
\end{equation}
However, the Lorentz factor must satisfy $\sLt\ge1$, which gives
the condition on the vielbein,
\begin{align}
\sLt^{-2} & =1-\sLv^{\prime}\sLv=1-e_{1}^{\tr}(e_{3}^{\tr}\sEta e_{3})^{-1}e_{1}\le1.\label{eq:llt-cond}
\end{align}
Hence, the vielbein can be triangularized by $\Lambda$ iff (\ref{eq:llt-cond})
holds.

On the other hand, the metric obtained from the vielbein $E$ has
the form, 
\begin{equation}
\gMet=E^{\tr}\eta E=\begin{pmatrix}-e_{0}e_{0}+e_{2}^{\tr}\sEta e_{2} & \, & -e_{0}e_{1}^{\tr}+e_{2}^{\tr}\sEta e_{3}\\
-e_{1}e_{0}+e_{3}^{\tr}\sEta e_{2} &  & -e_{1}e_{1}^{\tr}+e_{3}^{\tr}\sEta e_{3}
\end{pmatrix}\!.\label{eq:g-arb-E}
\end{equation}
The metric can be \nPlusOne{} decomposed (\ref{eq:g-N+1}) such that
$\gLapse^{2}>0$ (i.e., $\gLapse$ is real) and $\gSp$ is positive
definite if the foliation is spacelike. Combining (\ref{eq:g-arb-E})
with (\ref{eq:g-N+1}) yields,\bSe
\begin{align}
\gLapse^{2} & =\left(e_{0}-e_{1}^{\tr}e_{3}^{-1}e_{2}\right)^{2}\left(1-e_{1}^{\tr}(e_{3}^{\tr}\sEta e_{3})^{-1}e_{1}\right)^{-1},\label{eq:deriv-2}\\
\gSp & =e_{3}^{\tr}\left(\sEta-e_{3}^{-1,\tr}e_{1}e_{1}^{\tr}e_{3}^{-1}\right)e_{3}.
\end{align}
\eSe The spatial metric is positive definite $\|\gSp\|>0$ if $\|e_{3}^{-1,\tr}e_{1}e_{1}^{\tr}e_{3}^{-1}\|<1$,
and also $\gLapse^{2}>0$ if $1-e_{1}^{\tr}(e_{3}^{\tr}\sEta e_{3})^{-1}e_{1}\ge0$,
which is exactly the same as the condition (\ref{eq:llt-cond}).\hfill$\square$\bigskip{}

Now, assuming a common spacelike hypersurface for both the metrics
(which exists if and only if the principal square root exists \cite{Hassan:2017ugh}),
we can triangularize the vielbeins in (\ref{eq:sym-cond}) using the
above lemma. Then, starting from the vielbeins in the lower triangular
form,
\begin{equation}
E(\gMet)=\begin{pmatrix}\gLapse & 0\\
\gE\gShiftVec & \gE
\end{pmatrix}\!,\qquad E(\fMet)=\begin{pmatrix}\fLapse & 0\\
\bar{\fE}\fShiftVec & \bar{\fE}
\end{pmatrix}\!,
\end{equation}
the symmetrization condition (\ref{eq:sym-cond}) is equivalent to
$A=\eta\Lambda E(\fMet)E(\gMet){}^{-1}=A^{\tr}$ where,\bSe
\begin{gather}
A=\begin{pmatrix}-1 & 0\\
0 & \sEta
\end{pmatrix}\!\cdot\!\begin{pmatrix}\sLt\, & \sLp^{\prime}\\
\sLp & \sLs
\end{pmatrix}\!\!\begin{pmatrix}1 & 0\\
0 & \sRs
\end{pmatrix}\!\cdot\!\begin{pmatrix}\fLapse & 0\\
\bar{\fE}\fShiftVec & \bar{\fE}
\end{pmatrix}\!\cdot\!\begin{pmatrix}\gLapse^{-1} & 0\\
-\gShiftVec\gLapse^{-1} & \gE^{-1}
\end{pmatrix}\!,
\end{gather}
which can be expanded as (note $\sLs\sLp\sLtinv=\sLp$),
\begin{equation}
A=\begin{pmatrix}\,-\sLt\gLapse^{-1}-\sLp^{\prime}\sRs\bar{\fE}(\fShiftVec-\gShiftVec)\gLapse^{-1}\, & \, & -\sLp^{\prime}\sRs\bar{\fE}\gE^{-1}\,\\
\,\sEta\sLp\gLapse^{-1}+\sEta\sLs\sRs\bar{\fE}(\fShiftVec-\gShiftVec)\gLapse^{-1} &  & \sEta\sLs\sRs\bar{\fE}\gE^{-1}\,
\end{pmatrix}\!.
\end{equation}
\eSe Therefore, (\ref{eq:sym-cond}) is equivalent to $A=A^{\tr}$,
which reads,\bSe\label{eq:sym-cond-1}
\begin{align}
\gShiftVec-\gLapse\gE^{-1}\sLp\sLtinv & =\fShiftVec+\fLapse(\sRs\bar{\fE})^{-1}\sLp\sLtinv,\label{eq:sym-h-lapse}\\
\sEta\sLs\sRs\bar{\fE}\gE^{-1} & =\big(\sEta\sLs\sRs\bar{\fE}\gE^{-1}\big)^{\tr}.\label{eq:sym-h-sp}
\end{align}
\eSe In fact, the equation (\ref{eq:sym-h-lapse}) is the shift vector
$\hShiftVec$, and (\ref{eq:sym-h-sp}) is the spatial part of $\hMet$
(\ref{eq:hSp}), so the parametrization from section \ref{ssec:param}
satisfies (\ref{eq:sym-cond-1}). Since $\sLp$ and $\hShiftVec$
are arbitrary vector fields, the parametrization is exhausting, generating
all possible $S=E(\gMet)^{-1}\eta\Lambda E(\fMet)$.

\section{Components of $V_{g}$ and $V_{f}$}

\label{app:VgVf_components}

The components of $\gVse$ can be determined using (\ref{eq:proj-X}),\bSe
\begin{equation}
\gLapse\tud{\gVse}00=-\gLapse\grho-\gShift^{i}\gjota_{i},\qquad\gLapse\tud{\gVse}0i=-\gjota_{i},\qquad\gLapse\tud{\gVse}ik=\gShift\tud{\gJota}ik+\gShift^{i}\gjota_{k},
\end{equation}
\eSe where $\tud{\gVse}i0=\grho\gShift^{i}-\gLapse\gjota^{i}+\tud{\gVse}ik\gShift^{k}$.
Then,\bSe
\begin{align}
\gLapse\tud{\gVse}00 & =\gLapse\,e_{n}(\sfD),\qquad\gLapse\tud{\gVse}0i=\gSp_{ij}\big\llbracket\sgB\,Y_{n-1}(\sfD)\tud{\big\rrbracket}jk\sgn^{k},\\
\gLapse\tud{\gVse}ik & =\gLapse\,e_{n}(\sfD)\tud{\delta}ik-\gLapse\big\llbracket\sfD\,Y_{n-1}(\sfD)\tud{\big\rrbracket}ik+\fLapse\big\llbracket\sLtinv Y_{n-1}(\sgB)\tud{\big\rrbracket}ik.
\end{align}
\eSe Note that (\ref{eq:id-alg2}) can be expanded,\bSe
\begin{align}
\gLapse\det\gE\,\tud{\gVse}00+\fLapse\det\fE\,\tud{\fVse}00 & =\gLapse\det\gE\,V,\\
\gLapse\det\gE\,\tud{\gVse}0i+\fLapse\det\fE\,\tud{\fVse}0i & =0,\\
\gLapse\det\gE\,\tud{\gVse}ik+\fLapse\det\fE\,\tud{\fVse}ik & =\gLapse\det\gE\,V\delta_{k}^{i},
\end{align}
\eSe where $V=e_{n}(S)=e_{n}(\sfD)+\gLapse^{-1}\fLapse\,\sLtinv e_{n-1}(\sgB)=\sgV+\gLapse^{-1}\fLapse\sfV$.
Hence,\bSe
\begin{align}
\fLapse\frac{\det\fE}{\det\gE}\,\tud{\fVse}00 & =\fLapse\,\sLtinv e_{n-1}(\sgB),\\
\fLapse\frac{\det\fE}{\det\gE}\,\tud{\fVse}0i & =-\gSp_{ij}\tud{\big\llbracket\,\sgB\,Y_{n-1}(\sfD)\,\big\rrbracket}jk\sgn^{k},\\
\fLapse\frac{\det\fE}{\det\gE}\,\tud{\fVse}ik & =\fLapse\,\sLtinv e_{n-1}(\sgB)\tud{\delta}ik+\gLapse\big\llbracket\tud{\sfD\,Y_{n-1}(\sfD)\,\big\rrbracket}ik-\fLapse\tud{\big\llbracket\sLtinv Y_{n-1}(\sgB)\,\big\rrbracket}ik.
\end{align}
\eSe Finally, the effective stress-energy tensors reads,\bSe
\begin{alignat}{2}
\gLapse\,\tud{\gVse}00 & =\gLapse\sgV, & \fLapse\frac{\det\fE}{\det\gE}\,\tud{\fVse}00 & =\fLapse\sfV,\\
\gLapse\,\tud{\gVse}0i & =\gSp_{ij}\tud{\sgQU}jk\sgn^{k}, & \fLapse\frac{\det\fE}{\det\gE}\,\tud{\fVse}0i & =-\gSp_{ij}\tud{\sgQU}jk\sgn^{k},\\
\gLapse\,\tud{\gVse}ik & =\gLapse\sgV\delta_{k}^{i}-\gLapse\tud{\sfU}ik+\fLapse\tud{\sgU}ik, & \qquad\fLapse\frac{\det\fE}{\det\gE}\,\tud{\fVse}ik & =\fLapse\sfV\delta_{k}^{i}+\gLapse\tud{\sfU}ik-\fLapse\tud{\sgU}ik.
\end{alignat}
\eSe

\section{Projections of $\nabla_{\mu}\protect\tud{V_{g}}{\mu}{\nu}=0$}

\label{app:proj-nabla-Vg}

From (\ref{eq:var-en-X}) follows (valid for any $X$),\vspace{-1ex}
\begin{equation}
\delta e_{n}(X)=\Tr\left[Y_{n-1}(X)\,\delta X\right]=\frac{1}{2}\Tr\left[Y_{n-1}(X)\,X^{-1}\delta X^{2}\right].
\end{equation}
Then for $\delta\sgV=\delta e_{n}(\sfD)$ we have,
\begin{align}
\delta\sgV & =\frac{1}{2}\Tr\left[\sfD^{-1}Y_{n-1}(\sfD)\,\delta\sfD^{2}\right]=\frac{1}{2}\Tr\left[\sfD^{-1}Y_{n-1}(\sfD)\,\delta(\gSp^{-1}\fSp-\sgn\sgn^{\tr}\fSp)\right]\\
 & =\frac{1}{2}\Tr\Big[\gSp\sgQU\,\delta\gSp^{-1}+\sfU\fSp^{-1}\delta\fSp-\sgn^{\tr}\gSp\sgQU\,\delta\sgn-\gSp\sgQU\sgn\,\delta\sgn^{\tr}\Big].
\end{align}
Since $\Tr\left[\sgn^{\tr}\gSp\sgQU\,\delta\sgn\right]=\Tr\left[\gSp\sgQU\sgn\,\delta\sgn^{\tr}\right]=-\gjota_{k}\delta\sgn^{k}$,
we conclude, 
\begin{align}
\delta\sgV & =\frac{1}{2}\sgQU_{ij}\delta\gSp^{ij}+\frac{1}{2}\sfU^{ij}\delta\fSp_{ij}+\gjota_{k}\delta\sgn^{k}.
\end{align}
Here, the raising and lowering indices is done with the respective
spatial metric. From $\grho+\sgV=\gjota_{k}\sgn^{k}$ follows immediately,
\begin{equation}
\delta\grho=\frac{1}{2}\sgQU^{ij}\delta\gSp_{ij}-\frac{1}{2}\sfU^{ij}\delta\fSp_{ij}+\sgn^{k}\delta j_{k}.
\end{equation}
Expanding the variation $\delta$ as $\partial_{\mu}$ in terms of
$\partial_{\mu}\gSp_{ij}$, $\partial_{\mu}\fSp_{ij}$, and $\partial_{\mu}\sgn^{i}$
yields (\ref{eq:lem1a}). This proves the first half of Lemma~\ref{lemma:vgf-flux}.
Now, substitute the following relations back into (\ref{eq:lem1a-1}),\bSe\label{eq:apx-met-der}
\begin{align}
\frac{1}{2}\sgQU^{ij}\partial_{t}\gSp_{ij} & =\tud{\sgQU}ij\big(\signK\gLapse\tud{\gK}ji+\gD_{i}\gShift^{j}\big), & \frac{1}{2}\sgQU^{ij}\partial_{k}\gSp_{ij} & =\tud{\sgQU}ij\tud{\gCS}j{ik},\\
\frac{1}{2}\sfU^{ij}\partial_{t}\fSp_{ij} & =\tud{\sfU}ij\big(\signK\fLapse\tud{\fK}ji+\fD_{i}\fShift^{j}\big), & \frac{1}{2}\sfU^{ij}\partial_{k}\fSp_{ij} & =\tud{\sfU}ij\tud{\fCS}j{ik}.
\end{align}
\eSe We get,\vspace{-2ex}
\begin{align}
\qvf^{k}\partial_{\qvf}\grho-\sgn^{i}\qvf^{k}\partial_{\qvf}\gjota_{i} & =\tud{\sgQU}ij\tud{\gCS}j{ik}\qvf^{k}-\tud{\sfU}ij\tud{\fCS}j{ik}\qvf^{k}\\
 & =\tud{\sgQU}ij\left(\gD_{i}-\partial_{i}\right)\qvf^{j}-\tud{\sfU}ij\left(\fD_{i}-\partial_{i}\right)\qvf^{j}.
\end{align}
Moving the derivatives to the left then using $\tud{\sfU}ij=\sgn^{i}\gjota_{j}+\tud{\sgQU}ij$,
$\gD_{k}\gjota_{i}=\partial_{k}\gjota_{i}-\tud{\gCS}j{ik}\gjota_{j}$,
and $\gD_{i}\qvf^{j}=\partial_{i}\qvf^{j}+\tud{\gCS}j{ik}\qvf^{k}$,
gives,
\begin{equation}
\tud{\sgQU}ij\gD_{i}\qvf^{j}-\tud{\sfU}ij\fD_{i}\qvf^{j}=\qvf^{k}\gD_{k}\grho-\sgn^{i}\qvf^{k}\partial_{k}\gjota_{i}+\left(\tud{\sgQU}ij-\tud{\sfU}ij\right)\partial_{i}\qvf^{j}=\Lie_{\qvf}\grho-\sgn^{i}\Lie_{\qvf}\gjota_{i}.
\end{equation}
This proves (\ref{eq:lem1b-1}). The identity (\ref{eq:lem1b-2})
follows from $\Lie_{\qvf}\grho-\sgn^{i}\Lie_{\qvf}\gjota_{i}=-\Lie_{\qvf}\sgV+\gjota_{i}\Lie_{\qvf}\sgn^{i}$.

Now we turn our attention to the projections of $\gCD_{\mu}\tud{\gVse}{\mu}{\nu}=0$.
Plugging in $\partial_{t}\gSp_{ij}$ and $\partial_{t}\fSp_{ij}$
from (\ref{eq:apx-met-der}) into (\ref{eq:lem1a-1}), as well as
$\partial_{t}\grho$ and $\partial_{t}\gjota_{i}$ from (\ref{eq:proj-sv}),
we obtain,
\begin{align}
0 & =\Lie_{\gShiftVec}\grho-\gLapse\gD_{i}\gjota^{i}-2\gjota^{i}\gD_{i}\gLapse\isignK\gLapse\gK\grho\isignK\gLapse\tud{\gJota}ij\tud{\gK}ji,\nonumber \\
 & \qquad-\,\sgn^{i}\left[\Lie_{\gShiftVec}\gjota_{i}-\gD_{j}\left(\gLapse\tud{\gJota}ji\right)-\grho\gD_{i}\gLapse\isignK\gLapse\gK\gjota_{i}\right]\nonumber \\
 & \qquad-\,\sgQU^{ij}\Big(\usignK\gLapse\gK_{ij}+\gD_{i}\gShift_{j}\Big)+\sfU^{ij}\Big(\usignK\fLapse\fK_{ij}+\fD_{i}\fShift_{j}\Big).
\end{align}
Substituting $\gLapse\tud{\gJota}ij=\gLapse\big[\sgV\tud{\delta}ij-\tud{\sgQU}ij\big]+\fLapse\tud{\sgU}ij$,
yields,
\begin{align}
0 & =\Lie_{\gShiftVec}\grho-\gLapse\gD_{i}\gjota^{i}-2\gjota^{i}\gD_{i}\gLapse\isignK\gLapse\gK\grho\isignK\gLapse\left(\sgV\tud{\delta}ij-\tud{\sgQU}ij\right)\tud{\gK}ji\isignK\fLapse\tud{\sgU}ij\tud{\gK}ji\nonumber \\
 & \qquad-\,\sgn^{i}\left[\Lie_{\gShiftVec}\gjota_{i}-\gD_{j}\left(\gLapse\sgV\tud{\delta}ji-\gLapse\tud{\sgQU}ji+\fLapse\tud{\sgU}ji\right)-\grho\gD_{i}\gLapse\isignK\gLapse\gK\gjota_{i}\right]\nonumber \\
 & \qquad\isignK\gLapse\tud{\sgQU}ij\tud{\gK}ji-\tud{\sgQU}ij\gD_{i}\gShift^{j}\signK\fLapse\tud{\sfU}ij\tud{\fK}ji+\tud{\sfU}ij\fD_{i}\fShift^{j}.
\end{align}
Using (\ref{eq:lem1b-2}) from the lemma, then using $\grho+\sgV=\gjota_{k}\sgn^{k}$
where $\gjota_{i}=-\sgQU_{ik}\sgn^{k}$, we have,
\begin{align}
0 & =\tud{\sgQU}ij\gD_{i}\gShift^{j}-\tud{\sfU}ij\fD_{i}\gShift^{j}-\gLapse\gD_{i}\gjota^{i}-2\gjota^{i}\gD_{i}\gLapse\isignK\gLapse\gK\left(\grho+\sgV\right)\nonumber \\
 & \qquad+\,\sgn^{i}\gD_{i}\left(\gLapse\sgV\right)-\sgn^{i}\gD_{j}\left(\gLapse\tud{\sgQU}ji\right)+\sgn^{i}\gD_{j}\left(\fLapse\tud{\sgU}ji\right)+\grho\sgn^{i}\gD_{i}\gLapse\signK\gLapse\gK\,\sgn^{i}\gjota_{i}\nonumber \\
 & \qquad-\tud{\sgQU}ij\gD_{i}\gShift^{j}+\tud{\sfU}ij\fD_{i}\fShift^{j}\isignK\fLapse\tud{\sgU}ij\tud{\gK}ji\signK\fLapse\tud{\sfU}ij\tud{\fK}ji.\label{eq:apx-NM0}
\end{align}
Expanding the covariant derivatives in (\ref{eq:apx-NM0}) then canceling
terms gives,
\begin{align}
0 & =\gLapse\Big[-\tud{\sfU}ij\fD_{i}\sgn^{j}+\tud{\sgQU}ij\gD_{i}\sgn^{j}+\sgn^{i}\gD_{i}\sgV\Big]\nonumber \\
 & \qquad+\,\fLapse\Big[\uisignK\tud{\sgU}ij\tud{\gK}ji\signK\tud{\sfU}ij\tud{\fK}ji+\sgn^{i}\gD_{i}\tud{\sgU}ij-\tud{\sfU}ij\fD_{i}\sfn^{j}\Big].\label{eq:apx-NM1}
\end{align}
The factor of $\gLapse$ in (\ref{eq:apx-NM1}) vanishes identically;
this is again a consequence of the lemma noting that $j_{i}\Lie_{\sgn}\sgn^{i}=0$.
Therefore,
\begin{equation}
\tud{\sgU}ij\gD_{i}\sgn^{j}\signK\tud{\sgU}ij\tud{\gK}ji+\tud{\sfU}ij\fD_{i}\sfn^{j}\isignK\tud{\sfU}ij\tud{\fK}ji=\gD_{i}\big[\tud{\sgU}ij\sgn^{j}\big].\label{eq:apx-constr}
\end{equation}

To rewrite the above equation in a more symmetric form (exhibiting
the $\gMet\leftrightarrow\fMet$ duality) we note that for any $\tud X{\mu}{\nu}$
and $\varepsilon^{\mu}$ holds the identity,
\begin{equation}
\gCD_{\mu}\big(\tud X{\mu}{\nu}\varepsilon^{\nu}\big)-\fCD_{\mu}\big(\tud X{\mu}{\nu}\varepsilon^{\nu}\big)=-\tud X{\mu}{\nu}\varepsilon^{\nu}\partial_{\mu}\log\frac{\sqrt{-\fMet}}{\sqrt{-\gMet}}.
\end{equation}
A similar expression can be stated on the spatial slice, yielding,
\begin{equation}
\gD_{i}\big(\tud{\sfU}ij\sfn^{j}\big)-\fD_{i}\big(\tud{\sfU}ij\sfn^{j}\big)=\frac{\partial_{i}\sqrt{\gSp}}{\sqrt{\gSp}}\tud{\sfU}ij\sfn^{j}-\frac{\partial_{i}\sqrt{\fSp}}{\sqrt{\fSp}}\tud{\sfU}ij\sfn^{j}.\label{eq:cov-diff}
\end{equation}
Substituting (\ref{eq:cov-diff}) into (\ref{eq:apx-constr}) gives,
\begin{align}
\frac{1}{2}\gD_{i}\big(\tud{\sgU}ij\sgn^{j}\big)+\frac{1}{2}\fD_{i}\big(\tud{\sfU}ij\sfn^{j}\big) & =\tud{\sgU}ij\Bigg[\gD_{i}\sgn^{j}-\frac{1}{2}\frac{\partial_{i}\sqrt{\gSp}}{\sqrt{\gSp}}\sgn^{j}\signK\tud{\gK}ji\Bigg]\nonumber \\
 & \qquad+\,\tud{\sfU}ij\Bigg[\fD_{i}\sfn^{j}+\frac{1}{2}\frac{\partial_{i}\sqrt{\fSp}}{\sqrt{\fSp}}\sfn^{j}\isignK\tud{\fK}ji\Bigg]\!.\label{eq:proj-constr-sym}
\end{align}
All the covariant derivatives in (\ref{eq:proj-constr-sym}) can be
easily converted to partial derivatives of the respective metric since
$\gSp\sgU$ and $\fSp\sfU$ are symmetric.

\section{The notational Rosetta Stone}

\label{app:rosetta}

Table \ref{tab:roseta} shows the relation to the variables from the
papers of Hassan \& Rosen (HR) \cite{Hassan:2011zd,Hassan:2011ea}
and Hassan \& Lundkvist (HL) \cite{Hassan:2018mbl}.

\begin{table}[H]
\caption{\label{tab:roseta}Notational comparison to literature.}
\vspace{1ex}

\bgroup\centering\renewcommand{\arraystretch}{1.4}
\noindent \begin{centering}
\begin{tabular}{|c|c|l|l|}
\hline 
HR/HL & \multicolumn{2}{c|}{This work} & Comments\tabularnewline
\hline 
$\hrColor{\gamma}$ & $\gSp$ & $=\gE^{{\scriptscriptstyle \tr}}\sEta\gE$ & The spatial projection of $\gMet$\tabularnewline
\hline 
$^{3}\!f$ or $\phi$ & $\fSp$ & $=\fE^{{\scriptscriptstyle \tr}}\sEta\fE$ & The spatial projection of $\fMet$\tabularnewline
\hline 
$\hrD\hrn$ & $\sgn$ & $=\gE^{-1}\sLv$ & The boost parameter as seen by $\gMet$\tabularnewline
\hline 
$\hrn$ & $\sfn$ & $=\fE^{-1}\sLv$ & The boost parameter as seen by $\fMet$\tabularnewline
\hline 
$\hrQ/\hrx$ & $\sfQ$ & $=\fE^{-1}\sLs^{2}\fE$ & The spatial part of the boost squared\tabularnewline
\hline 
$\hrx^{-1}$ & $\sLt^{2}$ & $=1+\sLp^{{\scriptscriptstyle \tr}}\sEta\sLp$ & The Lorentz factor squared ($\sLp=\sLt\sLv$)\tabularnewline
\hline 
$\sqrt{\hrx}$ & $\sLtinv$ & $=(1-\sLv^{{\scriptscriptstyle \tr}}\sEta\sLv)^{1/2}$ & The reciprocal Lorentz factor\tabularnewline
\hline 
$\hrD\hrQ\hrD$ & $\gSp^{-1}\fSp$ & $=\sgD^{-1}\sfD=\sgB\sfD$ & The spatial square root squared\tabularnewline
\hline 
$\hrD\hrQ/\sqrt{\hrx}$ & $\sgB$ & $=\gE^{-1}\sLs\fE=\sgD^{-1}$ & 1st half of $\gSp^{-1}\fSp=\sgD^{-1}\sfD$\tabularnewline
\hline 
$\sqrt{\hrx}\hrD$ & $\sfD$ & $=\gE^{-1}\sLs^{-1}\fE$ & 2nd half of $\gSp^{-1}\fSp=\sgD^{-1}\sfD$\tabularnewline
\hline 
$\hrV$ & $\sgV$ & $=\betaSum\,e_{n}(\sfD)$ & $\gLapse e_{n}(S)=\gLapse\sgV+\fLapse\sfV$\tabularnewline
\hline 
$\tud{\hrU}ij$ & $\sgU$ & $=\betaSum\,\sLtinv Y_{n-1}(\sgB)$ & Symmetric with respect to $\gSp$\tabularnewline
\hline 
$\tud{\hrVbar}ij$ & $\sgQU$ & $=\betaSum\,\sgB\,Y_{n-1}(\sfD)$ & Symmetric with respect to $\gSp$\tabularnewline
\hline 
$\tud{\hrWbar}ij$ & $\sfU$ & $=\betaSum\,\sfD\,Y_{n-1}(\sfD)$ & Symmetric with respect to $\fSp$\tabularnewline
\hline 
$\tud{\hrSV}ij$ &  & $=\sgV\sI-\sgQU$ & Symmetric with respect to $\gSp$\tabularnewline
\hline 
$\tud{\hrUtilde}ij$ &  & $=\sfV\sI-\sfQU$ & Symmetric with respect to $\fSp$\tabularnewline
\hline 
\end{tabular}
\par\end{centering}
\egroup
\end{table}

\noindent The HR/HL variables are derived by varying the action in
3+1 form, while the variables in this work are obtained by projecting
the field equations. Based on Table~\ref{tab:roseta}, we conclude,
\begin{alignat}{1}
\hrV & =\betaSum\,e_{n}(\sqrt{x}\hrD),\\
\tud{\hrU}ij & =\betaSum\,\sqrt{x}\,Y_{n-1}(\sqrt{x}^{-1}\hrD\hrQ),\\
\tud{\hrVbar}ij & =\betaSum\,\sqrt{x}^{-1}\hrD\hrQ\,Y_{n-1}(\sqrt{x}\hrD),\\
\tud{\hrWbar}ij & =\betaSum\,\sqrt{x}\hrD\,Y_{n-1}(\sqrt{x}\hrD),\\
\tud{\hrUtilde}ij & =\betaSum\,\big[\sqrt{x}\,e_{n-1}(\sqrt{x}^{-1}\hrD\hrQ)-\sqrt{x}^{-1}\hrQ\,Y_{n-1}(\sqrt{x}^{-1}\hrD\hrQ)\big],
\end{alignat}
with the identity (\ref{eq:rho-v-identity}) expressed by,
\begin{equation}
e_{n}(\sqrt{x}^{-1}\hrD\hrQ)-e_{n}(\sqrt{x}\hrD)=(\hrDn)^{i}\hrVbar_{ij}(\hrDn)^{j}\,=\hrn^{i}\hrWbar_{ij}\hrn^{j}.
\end{equation}

\section{Spherically symmetric variables}

\label{app:ssym-vars}

Below are the \nPlusOne{} variables evaluated for the spherically
symmetric case,
\begin{align}
\gMet & =-\gAlpha^{2}\dd t^{2}+\gEA^{2}(\dd r+\gBeta\,\dd t)^{2}+\gEB^{2}(\dd\theta^{2}+\sin^{2}\theta\,\dd\phi^{2}),\\
\fMet & =-\fAlpha^{2}\dd t^{2}+\fEA^{2}(\dd r+\fBeta\,\dd t)^{2}+\fEB^{2}(\dd\theta^{2}+\sin^{2}\theta\,\dd\phi^{2}).
\end{align}
where $\gBeta=\hShift+\gAlpha\gEA^{-1}\sLp\sLtinv$ and $\fBeta=\hShift-\fAlpha\fEA^{-1}\sLp\sLtinv$.
Note that $\sLv=\sLp\sLtinv$ and $\sLt^{2}=1+\sLp^{2}$. The variables
instantiated from Table \ref{tab:vars} are,
\begin{alignat}{2}
\sgn^{r} & =\gEA^{-1}\sLp\sLtinv, & \sfn^{r} & =\fEA^{-1}\sLp\sLtinv,\\
\tud{\sgQ}rr & =\sLt^{2}, & \tud{\sfQ}rr & =\sLt^{2},\\
\tud{\sgQ}{\theta}{\theta} & =1, & \tud{\sfQ}{\theta}{\theta} & =1,\\
\tud{\sgD}rr & =\sLtinv\gEA\fEA^{-1}, & \tud{\sfD}rr & =\sLtinv\fEA\gEA^{-1},\\
\tud{\sgD}{\theta}{\theta} & =\gEB\fEB^{-1}=\sER^{-1}, & \tud{\sfD}{\theta}{\theta} & =\fEB\gEB^{-1}=\sER,\\
\tud{\sgB}rr & =\sLt\fEA\gEA^{-1}, & \tud{\sfB}rr & =\sLt\gEA\fEA^{-1},\\
\tud{\sgB}{\theta}{\theta} & =\fEB\gEB^{-1}=\sER, & \tud{\sfB}{\theta}{\theta} & =\gEB\fEB^{-1}=\sER^{-1},\\
\sgV & =\left\langle \sER\right\rangle _{0}^{2}+\sLtinv\fEA\gEA^{-1}\left\langle \sER\right\rangle _{1}^{2}, & \sfV & =\sLtinv\left\langle \sER\right\rangle _{1}^{2}+\fEA\gEA^{-1}\left\langle \sER\right\rangle _{2}^{2},\\
\tud{\sgU}rr & =\sLtinv\left\langle \sER\right\rangle _{1}^{2}, & \tud{\sfU}rr & =\sLtinv\fEA\gEA^{-1}\left\langle \sER\right\rangle _{1}^{2},\\
\tud{\sgU}{\theta}{\theta} & =\sLtinv\left\langle \sER\right\rangle _{1}^{1}+\fEA\gEA^{-1}\left\langle \sER\right\rangle _{2}^{1}, & \tud{\sfU}{\theta}{\theta} & =\sER\left\langle \sER\right\rangle _{1}^{1}+\sLtinv\fEA\gEA^{-1}\sER\left\langle \sER\right\rangle _{2}^{1},\\
\tud{\sgQU}rr & =\sLt\fEA\gEA^{-1}\left\langle \sER\right\rangle _{1}^{2}, & \tud{\sfQU}rr & =\sLt\left\langle \sER\right\rangle _{1}^{2},\\
\tud{\sgQU}{\theta}{\theta} & =\sER\left\langle \sER\right\rangle _{1}^{1}+\sLtinv\fEA\gEA^{-1}\sER\left\langle \sER\right\rangle _{2}^{1},\qquad & \tud{\sfQU}{\theta}{\theta} & =\sLtinv\left\langle \sER\right\rangle _{1}^{1}+\fEA\gEA^{-1}\left\langle \sER\right\rangle _{2}^{1},
\end{alignat}
where,
\begin{equation}
\left\langle \sER\right\rangle _{k}^{1}=\signV\betaScale\big(\beta_{k}+\beta_{k+1}\sER\big),\quad\left\langle \sER\right\rangle _{k}^{2}=\signV\betaScale\big(\beta_{k}+2\beta_{k+1}\sER+\beta_{k+2}\sER^{2}\big).
\end{equation}
All other components are zero. For any spatial operator $X$, we have
$\tud X{\theta}{\theta}=\tud X{\phi}{\phi}$. 

\clearpage